\documentclass{bmcart}

\RequirePackage{hyperref}
\usepackage{graphicx}
\usepackage{epsfig}
\usepackage{amsmath}
\graphicspath{{figures/}} % define the path to the folder with the figures
\DeclareMathOperator*{\argmax}{arg\,max}

\startlocaldefs
\endlocaldefs

\begin{document}

%%% Start of article front matter
\begin{frontmatter}

\begin{fmbox}
\dochead{Research}


\title{A note on the duality between interaction responses and mutual positions in flocking and schooling.}

%%%%%%%%%%%%%%%%%%%%%%%%%%%%%%%%%%%%%%%%%%%%%%
%%                                          %%
%% Enter the authors here                   %%
%%                                          %%
%% Specify information, if available,       %%
%% in the form:                             %%
%%   <key>={<id1>,<id2>}                    %%
%%   <key>=                                 %%
%% Comment or delete the keys which are     %%
%% not used. Repeat \author command as much %%
%% as required.                             %%
%%                                          %%
%%%%%%%%%%%%%%%%%%%%%%%%%%%%%%%%%%%%%%%%%%%%%%
\author[
   addressref={aff1},                   % id's of addresses, e.g. {aff1,aff2}
   corref={aff1},                       % id of corresponding address, if any
%   noteref={n1},                        % id's of article notes, if any
   email={andrea.perna@univ-paris-diderot.fr}   % email address
]{\inits{AP}\fnm{Andrea} \snm{Perna}}
\author[
   addressref={aff2},                   % id's of addresses, e.g. {aff1,aff2}
%   noteref={n1},                        % id's of article notes, if any
   email={guillaume.gregoire@univ-paris-diderot.fr}   % email address
]{\inits{GG}\fnm{Guillaume} \snm{Gregoire}}
\author[
   addressref={aff3},
   email={rmann@math.uu.se}
]{\inits{RPM}\fnm{Richard P} \snm{Mann}}

%%%%%%%%%%%%%%%%%%%%%%%%%%%%%%%%%%%%%%%%%%%%%%
%%                                          %%
%% Enter the authors' addresses here        %%
%%                                          %%
%% Repeat \address commands as much as      %%
%% required.                                %%
%%                                          %%
%%%%%%%%%%%%%%%%%%%%%%%%%%%%%%%%%%%%%%%%%%%%%%

\address[id=aff1]{%                           % unique id
  \orgname{Paris Interdisciplinary Energy Research Institute}, % university, etc
  \orgname{Paris Diderot University}, % university, etc
  \street{10 rue Alice Domon et Leonie Duquet},
  \postcode{75013},  % post or zip code
  \city{Paris}, % city
  \cny{France} % country
}
\address[id=aff2]{%                           % unique id
  \orgname{Laboratoire Matiere Systemes Complexes}, % university, etc
  \orgname{Paris Diderot University}, % university, etc
  \street{10 rue Alice Domon et L\'{e}onie Duquet},                     %
  \postcode{75013},  % post or zip code
  \city{Paris}, % city
  \cny{France}  % country
}
\address[id=aff3]{%
  \orgname{Mathematics Department},
  \orgname{Uppsala University},
  \street{L\"agerhyddsv\"agen 1},
  \postcode{75754},
  \city{Uppsala},
  \cny{Sweden}
}

%%%%%%%%%%%%%%%%%%%%%%%%%%%%%%%%%%%%%%%%%%%%%%
%%                                          %%
%% Enter short notes here                   %%
%%                                          %%
%% Short notes will be after addresses      %%
%% on first page.                           %%
%%                                          %%
%%%%%%%%%%%%%%%%%%%%%%%%%%%%%%%%%%%%%%%%%%%%%%

\begin{artnotes}
%\note{Sample of title note}     % note to the article
%\note[id=n1]{Equal contributor} % note, connected to author
\end{artnotes}

\end{fmbox}% comment this for two column layout

%%%%%%%%%%%%%%%%%%%%%%%%%%%%%%%%%%%%%%%%%%%%%%
%%                                          %%
%% The Abstract begins here                 %%
%%                                          %%
%% Please refer to the Instructions for     %%
%% authors on http://www.biomedcentral.com  %%
%% and include the section headings         %%
%% accordingly for your article type.       %%
%%                                          %%
%%%%%%%%%%%%%%%%%%%%%%%%%%%%%%%%%%%%%%%%%%%%%%

\begin{abstractbox}

\begin{abstract} % abstract
\parttitle{Background} 
Recent research in animal behaviour has contributed to determine how alignment, turning responses, and changes of speed mediate flocking and schooling interactions in different animal species. Here, we address specifically the problem of what interaction responses support different nearest neighbour configurations in terms of mutual position and distance. 
\parttitle{Results} %if any
We find that the different interaction rules observed in different animal species may be a simple consequence of the relative positions that individuals assume when they move together, and of the noise inherent with the movement of animals, or associated with tracking inaccuracy.
\parttitle{Conclusions} %if any
The anisotropic positioning of individuals with respect to their neighbours, in combination with noise, can explain several aspects of the movement responses observed in real animal groups, and should be considered explicitly in future models of flocking and schooling. By making a distinction between interaction responses involved in maintaining a preferred flock configuration, and interaction responses directed at changing it, we provide a frame to discriminate movement interactions that signal directional conflict from those underlying consensual group motion.
\end{abstract}

%%%%%%%%%%%%%%%%%%%%%%%%%%%%%%%%%%%%%%%%%%%%%%
%%                                          %%
%% The keywords begin here                  %%
%%                                          %%
%% Put each keyword in separate \kwd{}.     %%
%%                                          %%
%%%%%%%%%%%%%%%%%%%%%%%%%%%%%%%%%%%%%%%%%%%%%%


% MSC classifications codes, if any
%\begin{keyword}[class=AMS]
%\kwd[Primary ]{}
%\kwd{}
%\kwd[; secondary ]{}
%\end{keyword}

\end{abstractbox}
%
%\end{fmbox}% uncomment this for twcolumn layout

\end{frontmatter}

%%%%%%%%%%%%%%%%%%%%%%%%%%%%%%%%%%%%%%%%%%%%%%
%%                                          %%
%% The Main Body begins here                %%
%%                                          %%
%% Please refer to the instructions for     %%
%% authors on:                              %%
%% http://www.biomedcentral.com/info/authors%%
%% and include the section headings         %%
%% accordingly for your article type.       %%
%%                                          %%
%% See the Results and Discussion section   %%
%% for details on how to create sub-sections%%
%%                                          %%
%% use \cite{...} to cite references        %%
%%  \cite{koon} and                         %%
%%  \cite{oreg,khar,zvai,xjon,schn,pond}    %%
%%  \nocite{smith,marg,hunn,advi,koha,mouse}%%
%%                                          %%
%%%%%%%%%%%%%%%%%%%%%%%%%%%%%%%%%%%%%%%%%%%%%%

%%%%%%%%%%%%%%%%%%%%%%%%% start of article main body
% <put your article body there>

%%%%%%%%%%%%%%%%
%% Background %%
%%
\section*{Background}
Several animal species exhibit forms of collective motion in which two or more individuals move together coherently. Examples include flocks of migrating birds, schools of fish, murmurations of starlings, swarms of locusts, and many others. In general, the same group of animals can produce various types of collective patterns, including disordered aggregations, milling, or schooling depending on both internal states (e.g. hunger level) and external conditions (e.g. in response to a predator).

Much of our current understanding of collective motion of animal groups comes to us from the study of theoretical models, and in particular of a class of models known as `self-propelled particle models'. These models indicate that a small set of `rules' of interaction is sufficient to generate group level patterns that resemble, at least visually, with those formed by real animal groups. For instance, Reynolds~\cite{Reynolds} proposed a model that implements only three different rules. The first rule consists in a \textit{repulsion} behaviour, through which each individual turns away from its local neighbours and avoids local crowding and collisions. The second rule is an \textit{alignment} behaviour, or a turning response towards the average heading of local neighbours. The third rule is a turning response towards the position of more distant neighbours; this is an \textit{attraction} rule, in that it contributes to maintain the members of the group together. Several alternative models of collective motion have been proposed (see~\cite{Vicsek2012} for a review), each implementing a slightly different set of interaction rules. In spite of their differences, almost all the models existing in the literature are able to produce realistic looking patterns of collective behaviour, at least within a certain range of parameters.

The study of self-propelled particle models initially developed in fields outside biology, such as computer graphics and statistical physics, with the aim of understanding how coherent group behaviour emerges from local interactions. However, these models have since attracted increasing interest from biologists and researchers in animal behaviour, as a tool for addressing questions such as how individuals in a group `make decisions' together~\cite{Couzin2005,Conradt_et_al2009}, collectively avoid a predator~\cite{Ward2011}, or simply move together without a leader~\cite{Couzin2002}. In order to make meaningful predictions about the collective movement patterns of a given animal species, it is important that the interaction rules implemented in the models match those actually used by animals of that particular species. This exigence has pushed several research groups to collect empirical data on the movement of real animal groups, with the aim of validating the models.

There are two alternative ways to characterise the movement of animal groups: the first focuses on \textit{collective} behaviour, and consists in collecting data on the spatio-temporal organization of the group, such as e.g. the mutual positions of close neighbours; the second focuses instead on \textit{individual} behaviour. This latter approach operates by selecting a `focal individual' within the group, and recording all the changes of speed and direction of movement of that individual in response to the position and movement of its neighbours~\cite{Sumpter_Mann_Perna2012}. 

As an example of the first approach, Ballerini et al.~\cite{Ballerini2008} tracked the 3D positions of starlings flocking together in natural flocks, with the aim of characterising the spatial organization of the group. These authors observed that nearest neighbours consistently occupy the same positions with respect to each other, determining an anisotropic arrangement at the local scale. The anisotropy did not spread to the scale of the entire flock, but dropped quickly to a completely isotropic distribution between the sixth and the seventh nearest neighbour. The fact that the anisotropy cut-off depended on the number of neighbours, but not on the density of the group, was interpreted as evidence that starlings `pay attention' to a fixed `topological' number of six - seven neighbours, instead of responding to all neighbours within a fixed `metric' distance. A similar global level approach was adopted by Lukeman et al.~\cite{Lukeman2010}. These authors recorded the positions and orientations of surf scoters sitting on the water surface. The observed arrangements of neighbours around a focal individual were consistent with models implementing repulsion, alignment, and attraction, but also required the existence of a more direct interaction with one single neighbour situated in front. Buhl et al.~\cite{Buhl2012} measured the relative positions of swarming locusts, and observed isotropy in the radial distribution of neighbours around a focal individual. This distribution was compatible with both metric and topological models of interactions, but not with a third class of `pursuit/escape' models~\cite{Romanczuk2009} in which individuals try to reach neighbours ahead of and moving away from them, while they escape from other individuals that approach them from behind. Hemelrijk et al~\cite{Hemelrijk2010} measured how the overall shape (length \textit{vs.} width) of schools of mullets scales with group size. Their empirical data were consistent with a model in which the oblong shape of some schools results from individuals slowing down to avoid collisions.

Other studies have investigated the phenomenon of collective motion from the local level, by quantifying the responses of a focal individual to the movement of its conspecifics. For instance, Katz et al.~\cite{Katz2011} reconstructed the `force maps' that describe the acceleration and turning of schooling golden shiners, and Herbert-Read et al.~\cite{Herbert-Read2011} reconstructed the force maps of mosquitofish. These studies indicated that a fundamental component of how fish of both species interact are changes of speed: the fish consistently increased or decreased their speed to catch neighbours that they had respectively in front or behind; but when a neighbour was too close by, the speed responses were reversed, so speed changes also mediated collision avoidance. Both studies found only weak alignment responses, in comparison to attraction and repulsion forces. While both mosquitofish and golden shiners formed aligned groups, this was more a consequence of the fish following each other (and eventually becoming aligned) than an explicit alignment response. 

More recently, Pettit et al.~\cite{Pettit2013} applied a similar approach to the study of flight interactions in pigeons. The observed flocking responses of pigeons where different from those found in fish: alignment responses were explicit and strong, and collision avoidance was mainly mediated by turning, while speed remained relatively constant. These observations could be interpreted in terms of the different needs and constraints associated with flocking, which are different from those experienced by fish during schooling. Explicit alignment responses, for instance, might be necessary to achieve the high cohesion of pigeon flocks, that can fly without splitting for several kilometers. Avoiding collisions by turning away from the neighbour, instead of slowing down, might respond to a necessity to maintain a relatively constant speed, associated to the energetic constraints of flying.

If we focus on the spatial organization of pigeon flocks and fish schools, we can observe that pigeons were found to fly side by side most of the time, while both mosquitofish and golden shiners tended to have their closest neighbours directly in front or behind. We might speculate that the different relative positioning in these species is a direct result of the different interaction rules. Consider the case of an animal that avoids collisions by changing speed (like mosquitofish). Its acceleration response will be positive when the neighbour is in front and negative when the neighbour is behind, but will invert sign for close neighbours within the repulsion zone. There will be no acceleration response when the neighbour is on the border between attraction and repulsion zone. If turning does not mediate collision avoidance, turning response will be simply directed towards the neighbour, that is, the focal individual will turn to the left if its neighbour is on the left and will turn to the right if its neighbour is on the right. Only  neighbours that are exactly in front or exactly behind the focal individuals will not elicit any turning response. The positions at which mosquitofish are more likely to have their neighbours, that are directly in front and behind, hence correspond to those at which both the acceleration and turning responses are zero, at least when the focal individual and its neighbour are aligned. Similar arguments can be used to explain that when collision avoidance is mediated through turning away from the neighbour, but not through changing speed, a side by side configuration is the one for which both turning and acceleration responses are null. In other words, different interaction rules lead naturally to different local arrangements of neighbours within the group.

In the present paper, we examine the different implications of this duality between interaction rules and mutual positions in flocks and schools. We do so by focusing on the theoretical example of particles moving along the same trajectory at a fixed relative position from each other, either side by side, or in a front-back configuration. We model the imperfect ability of particles to stick to their target relative position, and the incertitude on position associated with tracking, by applying time-correlated, random displacements around the position of each particle.
Our analysis of these artificially generated trajectories addresses the question of what `apparent' response rules can be observed as a mere consequence of the imposed mutual positions and noise.

\section*{Results}

The movement of a focal individual with respect to a neighbour can be decomposed into an alignment response and an attraction-repulsion response by projecting it onto two different vectors (see figure~\ref{fig:illustrated_interactions}). Alignment is the component of movement response that has the same bearing as the neighbour. Attraction and repulsion correspond to the projection of focal individual's movement on the vector oriented towards its neighbour's body. In general, these two vectors are not orthogonal, except in very specific situations, such as when the focal individual and the neighbour move side by side in the same direction. In the extreme case when the focal individual and the neighbour are one behind the other, the alignment and the attraction/repulsion vectors coincide.

If the focal individual aims at keeping a fixed `target position' relative to its neighbour, for instance on its side, or behind it, we can imagine that it will spend most of the time in the proximity of that position, repeatedly moving away from it under the effect of noise, and actively heading back to it. Movements away from the target position, or back to it, can correspond to real animal movements, but can also result from noise associated with recording the position of the focal individual, such as GPS inaccuracy (in case of GPS tracking), or segmentation variability and pixelization (in case of video tracking).

Figure~\ref{fig:illustrated_interactions}-(a) shows a specific example with one individual, in red, having a preference for being directly behind its neighbour (target position marked by a star). A turn in the direction of the target position will be interpreted as an attraction (or repulsion) response; conversely, an alignment response would require to keep a straight direction, but this is not compatible with approaching the target. In figure~\ref{fig:illustrated_interactions}-(b), the relative positions of the focal individual and of its neighbour are the same, but the focal individual aims at reaching a schooling configuration side by side with its neighbour. The corresponding movement would be described in terms of an alignment response (the focal individual remains parallel to its neighbour), but also of attraction (because in this example reaching the target position involves getting closer to the neighbour). Both examples depict the same type of response (an attraction to the target), but we interpret them in terms of different alignment and attraction responses because we consider the other individual and not the target as the `point of attraction'.

The actual situation of two individuals moving together in two or three dimensions is more complicated, and involves not only different types of interactions e.g. alignment and/or attraction/repulsion, but also different types of responses, e.g. through turning, or acceleration, or both. In addition, in a real flocking situation individuals are not always aligned with each other and can have different speeds, making it more difficult to predict what interaction rules appear, on average, over a common trajectory. To test what interaction responses might support the movement of particles flocking together at a fixed distance and relative bearing, we simulate particles moving on the same trajectory but subject to small random displacements around these target positions (see methods). In particular, we focus on two configurations: one in which the two particles fly side by side, and one where the two particles fly one behind the other. Figure~\ref{fig:trajectory} illustrates one such generated trajectory for two particles moving side by side.

As expected, the side-by-side or front-back configurations imposed to the trajectories are reflected in the positions at which the neighbour is most frequently observed (figure~\ref{fig:apparent_rules_of_motion_of_particles_from_trajectory_analysis}-(a) and (d)). When the two trajectories are arranged in a front back configuration, the focal individual appears to turn in the direction of its neighbour with no `repulsion zone': independently of distance there is no zone in which turnings are directed away from the neighbour (figure~\ref{fig:apparent_rules_of_motion_of_particles_from_trajectory_analysis}-(b)). In this case, repulsion is mediated instead by changes of speed, as it is visible in figure~\ref{fig:apparent_rules_of_motion_of_particles_from_trajectory_analysis}-(c), where acceleration is positive for neighbours situated in front and negative for neighbours situated behind, but there is a region in which the polarity of the acceleration response is inverted, when the front-back distance to the neighbour is smaller than 5 m (the target distance between neighbouring particles implemented in the trajectories). These patterns of response are inverted for side by side trajectories: in this case, collision avoidance appears to be mediated through turning (figure~\ref{fig:apparent_rules_of_motion_of_particles_from_trajectory_analysis}-(e)), while changes of speed mediate attraction, but not collision avoidance (figure~\ref{fig:apparent_rules_of_motion_of_particles_from_trajectory_analysis}-(f)).

Our plots are similar to those obtained for real animal species, e.g. by Katz et al.~\cite{Katz2011} and Herbert-Read et al.~\cite{Herbert-Read2011} for fish moving prevalently in a front-back configuration and by Pettit et al.~\cite{Pettit2013} for pigeons flying side by side. The main difference is that in all studies on real animals, the repulsion zone had a roughly circular form, centered around the focal individual, while in out plots the repulsion zone has the form of a band, parallel or perpendicular to the direction of movement of the focal individual. This difference is likely due to the fact that in our trajectories, the target positions of the two particles are never exchanged for the entire duration of one ``flight'': one individual has its attractor always on the left side of its partner and the other individual always on the right side (or one individual always in front and the other always behind).  Real animals do switch from one to the other side of their neighbour (or from being in front to being behind), which means for instance that an animal situated roughly behind its neighbour ($\vartheta \simeq 0$ in figure \ref{fig:apparent_rules_of_motion_of_particles_from_trajectory_analysis}-(e)), and aiming at being on its side, will be nearly equally likely to turn left as to turn right, and on average will exhibit no consistent turning response.

Figure~\ref{fig:attraction_vs_alignment_of_particles_from_trajectory_analysis} plots the turning angle of the focal individual as a function of the direction of the neighbour (relative to the moving direction of the focal individual) and relative orientation. The figure is limited to the data points for which the focal individual has its neighbour in the attraction zone, i.e. when the mutual distance between the two individuals is larger than the average distance implemented in the trajectories (The Matlab\textsuperscript{\textregistered} code that we provide as electronic supplementary material has an easy to run interface to plot responses to neighbours in the attraction and repulsion zones, including acceleration responses and responses of individuals having different target positions). 

When the trajectories are arranged in a front-back configuration (figure~\ref{fig:attraction_vs_alignment_of_particles_from_trajectory_analysis}-(a)), the focal individual shows a strong turning response to face its neighbor's position, while alignment with the orientation of neighbors is not so much in evidence: the turning response in the figure is modulated along the $\theta$ axis, but presents almost no modulation along the $\phi$ axis. In the case of trajectories arranged side by side (figure~\ref{fig:attraction_vs_alignment_of_particles_from_trajectory_analysis}-(b)), the alignment response remains weak (modulation prevalently along the $\theta$ axis), but we also observe a collision avoidance response which depends on alignment: when the neighbour is in front and slightly on the left side of the focal individual ($\theta \simeq -\pi/6$), this latter turns to the right, and its response is stronger if the neighbour is also oriented to the right, i.e. in collision route with the focal individual.
It is interesting to observe how the attraction and alignment responses are altered when we increase the temporal autocorrelation of noise. A longer temporal autocorrelation of noise means that if, for example, an individual is on the left of the trajectory that it is supposed to follow, it will also remain on the left of the trajectory for longer time before returning back to the target position. Under these conditions, the plots of figure \ref{fig:attraction_vs_alignment_of_particles_from_trajectory_analysis}-(c) and (d) show a modulation along the alignment $\phi$ axis. In fact, with correlated noise the particles retain their component of movement parallel to the common trajectory, while their attraction to the target position is comparatively weaker.

A number of recent studies have quantified leadership in collectively moving groups by computing directional correlation delays~\cite{Nagy2010}. Directional correlation delays measure the average time delay within which one individual becomes aligned with a group neighbour, and it is assumed to indicate leadership behaviour if one individual consistently anticipates the direction taken by other members of the group. We computed directional correlation delays in our simulated data. When particles move side by side, there is no effect of being on the left or on the right, as we would have expected given the inherent left-right symmetry of the trajectories. When individuals move one behind the other, however, the individual in front appears to change direction first, and to be followed by its partner (see figure \ref{fig:directional_correlation_delay}). Intuitively we can see that when the common trajectory turns in one direction, the individual in front starts immediately turning in that direction, while the individual behind is projected temporarily to the opposite side of the curve. Increasing the temporal autocorrelation of noise does not change this, but it reduces the variability, because when errors on position are correlated, the estimation of direction of movement becomes more accurate.

By generating trajectories with three or more individuals at a fixed distance from each other, we can test the apparent responses to multiple neighbours. Even if in our simulations the three individuals do not respond to each other, but simply try each to keep a constant distance and orientation relative to the common trajectory, this does not prevent us from studying how \textit{apparent} responses to multiple neighbours are combined together. Figure~\ref{fig:multiple_interacting_neighbours} plots the observed acceleration (top row) and turning (bottom row) responses of a focal individual to two neighbours, for the case of three individuals moving in a front-back configuration. For this figure, the focal individual is randomly chosen between the three possible positions in the group (front, centre, back). The plots on the left in figure~\ref{fig:multiple_interacting_neighbours} report the average responses of the focal individual as a function of the front-back distance of the first and second neighbour; the plots on the right report the turning and acceleration responses that would be predicted by averaging pairwise interactions, that is, if the response of the focal individual resulted from the average of two independent interactions with individual neighbours as those presented in the top row of figure~\ref{fig:apparent_rules_of_motion_of_particles_from_trajectory_analysis} (for comparison with a similar analysis on real fish interactions see figure~3 of~\cite{Katz2011}). The combined responses to two neighbours are similar to those predicted from averaging pairwise interactions, but present larger modulations. This can be explained by considering that the position of all three individuals is affected by noise (or alternatively, that all three individuals can be randomly displaced by their target position). Hence, when the position of the focal individual appears to be displaced from its target relative to two neighbours, instead of just one, this provides increased evidence that the displacement is to be attributed to the focal individual, and not to the neighbours, and that the focal individual, and not one of the neighbours, is likely to show a compensatory response back to the target at the next time step.

\section*{Discussion}
% MUTUAL POSITIONS AS A RESULT OF RESPONSES
Several recent studies have mapped the `interaction rules' of flocking and schooling animals, expressed in terms of changes of speed and direction of movement in response to the position and movement of other individuals. It is well known that different interaction rules at the individual level produce different configurations at the group level. For example, different values of attraction and alignment are associated with a transition between a `liquid' configuration, in which individuals switch frequently their nearest neighbours, and a `solid' configuration, in which the positions of individuals are fixed relative to each other~\cite{Gregoire2003157}. Empirical studies have indicated that the strategy that animals adopt to avoid collisions affects the spatial positioning of nearest neighbours: animals that slow down to avoid collisions are more likely to occupy positions directly in front of their neighbours or directly behind them~\cite{Katz2011,Herbert-Read2011} and collision avoidance by changing speed is also responsible for the formation of elongated groups~\cite{Hemelrijk2010}; conversely, animals that turn away from their neighbours to avoid collisions are more likely to move in a side by side configuration~\cite{Pettit2013}. While we do not question the causal relation between interaction rules and configuration of the group, in the present study we revert it, to specifically explore how the relative positioning of individuals within a group depends on -and imposes individuals to adopt- different interaction rules.

% ASYMMETRIC POTENTIALS
The relative positioning of individuals, either side by side, or in a front-back configuration is sufficient to reproduce observed differences in the mechanisms used for collision avoidance, either by changing speed, or through turning. Anisotropic positioning of individuals with respect to their neighbours has been empirically observed in a number of species of collective moving animals, from fish~\cite{Partridge1980,Hemelrijk2010,Katz2011,Herbert-Read2011} to birds~\cite{Ballerini2008,Lukeman2010,Pettit2013} but it is not explicitly included into most self-propelled particle models of flocking and schooling. Some models involve a blind visual angle: a region of the visual field in which the presence of a neighbour does not induce any movement response (e.g. \cite{Couzin2002,Strombom:2011hc}), which can be considered as a form of anisotropy. However, these models otherwise consider attraction, alignment and repulsion as depending only on the distance from the neighbour, and not on its direction: interaction responses are organized in concentric regions around the focal individual. Outside animal behaviour, self-propelled particle models with anisotropic interaction zones have been studied in the context of collectively moving bacteria and other elongated or differently shaped particles (see e.g.~\cite{Wensink2014}). In these systems, the repulsion zone is determined directly by steric occlusions, and it typically leads to group formations organized in bands (smectic phases)~\cite{Wensink_and_Lowen2012}. In order to reproduce empirical observations, it seems important that future models of flocking and schooling take explicitly into account the anisotropy of interactions (it is bizarre how the empirical work of Ballerini and collaborators, one of the first detailed characterisations of anisotropic distribution of neighbours in flocks, triggered a large scientific debate about the topological - metric nature of interactions, but not about the anisotropy itself).

% FUNCTIONAL INTERPRETATION
While interaction responses and mutual positions are two complementary aspects of the same phenomenon, focusing on the interactions helps us understand the \textit{mechanisms} of flocking and schooling, while relative positions are more easily associated with the \textit{functions} of group movements. Addressing these different aspects together helps us reach a better understanding of flocking and schooling in relation to the biology of a species. For instance, if we focus on interaction rules, we can make the hypothesis that an animal that is unable to modulate quickly its speed -such as flying birds which might have difficulties to control independently speed and lift- will tend to use turning to avoid collisions, and this will lead it to form flocks with a side by side configuration. If instead we consider the side by side configuration as a target that animals aim to reach, and not as a by-product of interactions, we can argue that individuals moving side by side can both see each other, promoting bidirectional information transfer and collective decision making. We can also speculate about the energy efficiency of a side by side configuration, for instance whether it allows to take benefit from the vortices produced by the movement of neighbours.~\cite{Hemelrijk2014}

% NOISE
The interaction responses observed in our study can be interpreted in terms of animals constantly but imperfectly trying to keep an ideal mutual position. In theory, the same responses could also correspond to animals maintaining exactly the same `real' positions relative to each other (imagine the situation of two birds sitting on a boat), but whose `recorded' positions are affected by tracking noise. If the noise is uncorrelated, at each time step $t$ we expect to log a position for the focal individual that is displaced from its real position on average by the average absolute deviation of the noise distribution. In the case of Normally distributed noise with standard deviation $\sigma$ this average deviation is a simple proportion of the standard deviation: $\sigma \sqrt{\frac{2}{\pi}}$. Because the noise distribution at $t$ is uncorrelated with the distribution at $t-1$ and $t+1$, the particle at time $t$ will have just experienced -on average- an apparent movement directed from its real position at time $t-1$ to its recorded position at time $t$ of amplitude equal to the absolute deviation, and will -on average- experience the opposite movement from the recorded position to the real position between time $t$ and time $t+1$. In this extreme case, the observed interaction responses between neighbouring individuals can completely be described by this `regression to the mean' process, and the amplitude of `flocking responses' is in direct proportion to the standard deviation of the noise. Temporal correlation in the noise retards this regression to the mean, and appears in the plots as an alignment response, because in this case the movement of the focal individual remains parallel to the main trajectory of the pair in spite of its position being displaced inside the attraction or the repulsion zone. Autocorrelation in the noise can be introduced for instance by tracking algorithms that integrate prior expectations about the position of the target, which are likely to be implemented in many GPS and video tracking softwares. Because autocorrelations in the noise affect our ability to determine if flocking or schooling individuals exhibit alignment, it seems important that future studies try to estimate not only the amplitude of noise fluctuations, but also how these fluctuations are correlated in time.

%ENVIRONMENTAL TEMPLATES
In our simulations, individuals follow a pre-imposed trajectory, while keeping a constant relative position. Also real animals often follow `pre-imposed trajectories' in their collective movements. The simplest example are trails and zones clear of vegetation, but also conspicuous environmental features, such as the crest of a mountain and rivers can act as environmental templates that channel the movement of an animal group (e.g.~\cite{Mann06022011,Freeman23022011}). In laboratory experiments, the edges of the experimental setup also contribute to organize the movement of animals along preferential directions. Responses to neighbours and responses to environmental factors can be discriminated, provided that we can make realistic assumptions about these different interactions and how they are combined~\cite{bode2012distinguishing}). In practical situations, however, real animals can modulate the level of motivation, or intensity, with which they respond to their neighbours and to environmental targets, and for this reason it is not always possible to tell the contribution of these different factors apart. This problem is related to the problem of defining `leadership' with respect to route decision: if we consider the common route of a group as pre-imposed, then there is no leader within the group who decides what the route should be; if instead we consider that the group `builds its own route as it moves', then we can also ask what each member of the group contributes to the decision of this common route. We will describe soon how our analyses give us a hint about how to address these questions. 

% OUT-OF-EQUILIBRIUM INTERACTIONS
The flocking interactions observed in our study represent responses `at the equilibrium'. They describe the continuous adjustments that allow a flock or school to maintain a preferred configuration as the group moves. As such, they are not necessarily informative about when and how navigational decisions are taken: we would observe them even in the extreme case in which individuals have perfect agreement about the route to follow. Our simulations do actually imply such an agreement about a common route, in the sense that both particles follow the common trajectory with similar responses and no conflict. It is precisely in the presence of navigational conflict that we expect the equilibrium of mutual arrangements to be destabilized: interactions with environmental stimuli interfere with animal to animal interactions and induce individuals to abandon their mutual relative positions and relative alignment. This is in part captured by common measures of movement leadership such as the directional correlation delay~\cite{Nagy2010}, which implicitly assumes that leaders are those individuals that abandon more often their preferred orientation parallel to the neighbour, and followers are those individuals with a higher tendency to restore the aligned group configuration. In our analyses, directional correlation delays correlate with the position in front or on the back of the group. If we do not assume that trajectories are pre-imposed, but result from interactions, the individual that moves in front is also the first to draw the common trajectory, and it is reasonable to impute route decisions to this individual.

% LEADERSHIP
Leadership is also expressed by successful initiation of group movement, which also consists in one individual breaking the group configuration and other members of the group restoring it~\cite{King2010671,krause2000leadership}. Future studies should not be limited to characterise the average schooling and flocking responses of individuals, but should also focus on how deviations from, and returns to the group configuration `at the equilibrium' spread across the group. % ``Leaders'' need not necessarily be in front of the group: also an individual in the middle of the group can abandon its target position relative to others and initiate a movement in a new direction that propagates to the entire group. Stopping also implies abandoning a target position if other individuals are moving
Care should be taken, however, because changes in the internal configuration of a group do not exclusively reflect navigational conflict. For instance, when a group turns, the individuals on the larger radius face a conflict between speeding up, to maintain their position within the group, or abandoning their position, but avoid changing speed. Some positions within the group are also associated with hydrodynamic benefits that reduce energy expenditure~\cite{Killen22012012}. Navigational conflict and physiological constraints, such as a different ability of individuals to modulate speed and turning, interact in determining how the arrangement of neighbours within a group changes over time. For instance, in Pettit et al.~\cite{Pettit2013}, faster individuals were also more likely to get to the front of the group, and to become leaders in directional decisions when moving in group.

% NON-STEADY STATE MUTUAL POSITIONS
In addition to conflict about a common route to follow, the position itself to maintain relative to a neighbour can be at the origin of conflict. The simplest example is the case of one individual which wants to keep a certain distance $r$ from its partner, but the partner in turn aims at keeping a distance smaller or larger than $r$ from the first individual. The equilibrium configuration for the first individual does not correspond with the equilibrium configuration for the second individual, and vice-versa. A similar situation happens when the target positions are not symmetric, such as for instance if both individuals want to be directly in front of their neighbour, but not behind. In this case, changes of mutual position also reflect a conflict, which is not related to route decision, but to the relative position itself. A slightly more complicated example which can be described in terms of positional conflict are pursuit-escape situations, such as that of an individual chasing another individual. In such situations, even if the animals may appear to move together on a common trajectory, it is the mutual position, and not the trajectory itself that is at the origin of conflict. These possibilities should be taken into account when interpreting leadership measures such as directional (or speed) correlation delays: in the absence of positional conflict, an individual that abandons the ``equilibrium position'' is likely to be trading off its social needs (the need to have a neighbour at the preferred distance) and individual motivations (the attraction to an environmental feature), and correlation delays indicate a success in obtaining both group cohesion and movement towards the environmental target. Conversely, when positional conflict is present, a departure from the position that the group has can indicate leadership, if the change goes towards the preferred configuration of the focal individual, or followership if it goes towards the preferred configuration of its partner. 

% LARGER GROUPS
One of the open problems in research on collective motion is that of determining how individuals combine interactions with multiple neighbours. Here, we have shown that multiple neighbours can carry additional information about the movement of a focal individual not directly because they take part in the interactions, but indirectly because they reduce our uncertainty about the real position of the focal individual. If an animal group maintains a `solid-like' configuration, whereby individuals keep a constant position relative to their neighbours most of the time, like in our trajectories, the movement of a focal individual can be predicted in terms of its response to a single nearest neighbour, and including information about additional neighbours reduces uncertainty, but apart from this does not bring additional information. This might explain why information theoretical approaches, like the one adopted in~\cite{Herbert-Read2011} indicated that the movement of a focal individual can be predicted to a large amount by looking at only one nearest neighbour, and including further neighbours only marginally helped to improve the prediction. We are confident that future studies discriminating between interactions at the equilibrium and transient interactions will help to further improve our understanding of more complex patterns of response to multiple neighbours. 

\section*{Conclusion}
We have illustrated the duality between interaction rules and mutual positions in moving animal groups. This duality can be described in terms of two considerations. The first is that the neighbour-to-neighbour interactions that support collective motion are often anisotropic and lead to specific patterns of positioning of an animal relative to its neighbours. The second is that animals aim at keeping a particular position relative to their neighbours, and this can only be achieved by interaction responses with specific characteristics.

Our analyses suggest that movement interactions observed and quantified by recent studies on real animal group are largely determined by simple positional adjustments necessary to maintain a preferred local configuration of the group, and point to the necessity of discriminating between these interactions `at the equilibrium', and interactions that correspond to real navigational decisions.

\section*{Methods}
\subsection*{Trajectory generation}
We generated random trajectories, each having a length $N=2^{12}$ steps. The trajectories are defined by a sequence of step lengths (speed per time step) and a sequence of turnings intercalated between the steps.

The speed values $S$ are numbers extracted from the distribution
\begin{equation}
S = S_0 + s \frac{\epsilon_1(t)}{\max \left| \epsilon_1 \right|}
\end{equation}
and the turning angles $T$ are
\begin{equation}
T = a \frac{\epsilon_2(t)}{\max \left| \epsilon_2 \right|}
\end{equation}
In these equations, $\epsilon_1$ and $\epsilon_2$ represent sequences of temporally correlated random numbers and are generated as follows. We first generate $N$ random numbers uniformly distributed in the interval  $\left[ -0.5, 0.5 \right]$. In order to exclude abrupt changes of direction and speed, we apply to both sequences a low-pass temporal frequency filter with equation 
\begin{equation}
\epsilon(t) = \exp \left( - \frac{\omega(t)^2}{2\sigma^2} \right)
\label{eq:low-pass_filter}
\end{equation}
where $\omega$ are temporal frequencies and $\sigma$ controls the filter standard deviation. By setting $\sigma = \frac{N}{C_T}$, with a cut-off period for the temporal correlations $C_T = 300 \textrm{steps}$ we impose that speed and turning fluctuations typically occur over a period of 300 time steps, or longer. In our simulations, we fix arbitrarily $S_0 = 5$ and $s = 0.2$ metres per time step and $a = 0.02$ radians per time step. We further assume that 5 time steps in the trajectory correspond to one second of time. Our results are intended to illustrate qualitative differences in the observed patterns of movement, which remain stable for wide ranges of arbitrary parameters.
% Normalization of $\epsilon_1$ and $\epsilon_2$ by their maximum absolute value implies that the maximum of the distribution is fixed, but not the standard deviation or the mean.

The positions of individuals along the trajectory at time $t$ are determined by first drawing the segment that intersects the trajectory at $t$ and having a specific orientation $\theta$ relative to the segment of trajectory between $t$ and $t+1$, and selecting equally spaced points (at distance $r=5\textrm{m}$ from each other) on this segment. These individual trajectories represent the movement of an hypothetical focal individual and its partner (and in some simulations of a third individual) which successfully keep a constant distance and relative position to each other while moving together. % NOTE THAT THIS PROCESS DOES NOT CREATE IDENTICAL TRAJECTORIES, NOR TRAJECTORIES IN WHICH THE MEASURED BEARING IS IDENTICAL AT EVERY TIME STEP.
 
The `recorded' positions of the individuals do not match exactly those generated as above, but are displaced in a random direction at every time step, to simulate tracking noise, or an imperfect ability to maintain the desired flocking configuration. These displacements are autocorrelated in time, so that if an individual is for instance on the left of its target position at time $t$, it is more likely to be on the left of the target position also at time $t+1$. There is no cross-correlation between the random displacements of the focal individual and those of its neighbour. 
The random displacements are computed as follows. We first generate series of $N$ random numbers, normally distributed with mean 0 and standard deviation 1, then we apply a low-pass filter analogous to the one used in equation \ref{eq:low-pass_filter}, with cut-off frequency $\sigma_d = \frac{N}{C_D}$, where $C_D$ is the cut-off correlation period for displacements (the number of time steps after which the displacements become uncorrelated). In our simulations $C_D = 20 \textrm{steps}$ except when otherwise stated.
After the filtering operation, we rescale the numbers to obtain distributions with standard deviation $r/2$. Two random numbers taken from two such generated series describe the x and y components of the displacement. 

The analyses reported in the present manuscript focus on the comparison of two conditions. In the first condition the focal individual has a target position directly in front or behind its neighbour ($\theta = 0$) . In the second condition, the target position for the focal individual is on the side of its neighbour ($\theta = \pi/2$). For each condition, we generate 100 random trajectories. The order of individuals along the segment, that is, whether the focal individual is in front or behind its neighbour (respectively left or right when $\theta = \pi/2$) is constant for the whole length of one trajectory, but changes randomly from one trajectory to the other, with half of the trajectories on average displaying the focal individual on the left and the other half displaying it on the right. The movement responses observed in all trajectories are merged together for the analyses.

\subsection*{Data analysis}
At each time step $t$ we measure the instantaneous speed of the focal individual
$$s(t) = \sqrt{\left(x(t) -x(t-1) \right)^2 + \left(y(t) -y(t-1)\right)^2}/dt,$$ where $x(t)$ and $y(t)$ are the $x$ and $y$ coordinates of the focal individual at time $t$ and $dt$ is the duration of a time step. The direction of movement of the focal individual is
$$\psi(t) = \mathrm{atan2}\left(y(t)- y(t-1), x(t)- x(t-1)\right),$$.

The response of the focal individual to its neighbours is described by its tangential acceleration
$$a(t) = \left(s(t) - s(t-1)\right) / dt$$ and its speed of direction change
$$\alpha(t+1) = \left(\psi(t) - \psi(t-1)\right)/dt,$$ where care is taken to compute the correct angular difference, $\psi(t)-\psi(t-1)$, with regard to the periodicity of $\psi(t)$.

The relative position and orientation of a neighbour in the frame of reference of the focal individual are described by 
their observed mutual distance 
$$d_{ij} \left( t \right) = \sqrt{\left(x_j(t) -x_i(t) \right)^2 + \left(y_j(t) -y_i(t)\right)^2}$$, \\
and the direction $\theta$ of the neighbour in the frame of reference of the focal fish was\\
$$\vartheta_{ij}(t) =  \mathrm{atan2} \left( y_j(t) - y_i( t ), x_j( t )- x_i( t ) \right) - \alpha_i \left( t \right)$$.\\

The directional correlation delay $\tau^*$ is the time delay $\tau$ that maximizes the correlation of direction between the focal individual and its partner \\
$$ \tau_{ij}^* = \argmax_{\tau} \left\langle \cos \left( \psi_i(t) - \psi_j(t + \tau) \right) \right\rangle$$

The Matlab\textsuperscript{\textregistered} source code used to generate the trajectories and for all the analyses is available as online supplementary material.

%%%%%%%%%%%%%%%%%%%%%%%%%%%%%%%%%%%%%%%%%%%%%%
%%                                          %%
%% Backmatter begins here                   %%
%%                                          %%
%%%%%%%%%%%%%%%%%%%%%%%%%%%%%%%%%%%%%%%%%%%%%%

\begin{backmatter}

\section*{Competing interests}
The authors declare that they have no competing interests.

\section*{Author's contributions}
Designed research: AP RPM \\
Performed research: AP \\
Contributed ideas / analysis tools: GG \\
Wrote the paper: AP

\section*{Acknowledgements}
This work was supported by European Union Information and Communication Technologies project ASSISI\_bf 601074. AP was supported by the city of Paris (Research in Paris programme) and the Ile de France region (grant 01RA140024-RIDF-PERNA). The funders had no role in study design, data collection and analysis, decision to publish, or preparation of the manuscript.

%%  After submission of the .TEX file,                     %%
%%  you will be prompted to submit your .BBL file.         %%

% if your bibliography is in bibtex format, use those commands:
% \bibliographystyle{bmc-mathphys} % Style BST file
% \bibliography{/Users/perna/articoli/bibliografia}      % Bibliography file (usually '*.bib' )

% or include bibliography directly:

\newcommand{\BMCxmlcomment}[1]{}

\BMCxmlcomment{

<refgrp>

<bibl id="B1">
  <title><p>Flocks, herds and schools: A distributed behavioral
  model</p></title>
  <aug>
    <au><snm>Reynolds</snm><fnm>CW</fnm></au>
  </aug>
  <source>SIGGRAPH Comput. Graph.</source>
  <publisher>New York, NY, USA: ACM</publisher>
  <pubdate>1987</pubdate>
  <volume>21</volume>
  <issue>4</issue>
  <fpage>25</fpage>
  <lpage>-34</lpage>
</bibl>

<bibl id="B2">
  <title><p>Collective motion</p></title>
  <aug>
    <au><snm>Vicsek</snm><fnm>T</fnm></au>
    <au><snm>Zafeiris</snm><fnm>A</fnm></au>
  </aug>
  <source>Collective motion</source>
  <pubdate>2012</pubdate>
  <volume>517</volume>
  <fpage>71</fpage>
  <lpage>-140</lpage>
  <url>http://www.sciencedirect.com/science/article/pii/S0370157312000968</url>
</bibl>

<bibl id="B3">
  <title><p>Effective leadership and decision-making in animal groups on the
  move.</p></title>
  <aug>
    <au><snm>Couzin</snm><fnm>ID</fnm></au>
    <au><snm>Krause</snm><fnm>J</fnm></au>
    <au><snm>Franks</snm><fnm>NR</fnm></au>
    <au><snm>Levin</snm><fnm>SA</fnm></au>
  </aug>
  <source>Nature</source>
  <pubdate>2005</pubdate>
  <volume>433</volume>
  <issue>7025</issue>
  <fpage>513</fpage>
  <lpage>-516</lpage>
</bibl>

<bibl id="B4">
  <title><p>Leading According to Need in Self Organizing Groups.</p></title>
  <aug>
    <au><snm>Conradt</snm><fnm>L</fnm></au>
    <au><snm>Krause</snm><fnm>J</fnm></au>
    <au><snm>Couzin</snm><fnm>I. D.</fnm></au>
    <au><snm>Roper</snm><fnm>T. J.</fnm></au>
  </aug>
  <source>The American Naturalist</source>
  <publisher>The University of Chicago Press for The American Society of
  Naturalists</publisher>
  <pubdate>2009</pubdate>
  <volume>173</volume>
  <issue>3</issue>
  <fpage>pp.304</fpage>
  <lpage>312</lpage>
  <url>http://www.jstor.org/stable/10.1086/596532</url>
</bibl>

<bibl id="B5">
  <title><p>Fast and accurate decisions through collective vigilance in fish
  shoals.</p></title>
  <aug>
    <au><snm>Ward</snm><fnm>AJW</fnm></au>
    <au><snm>Herbert Read</snm><fnm>JE</fnm></au>
    <au><snm>Sumpter</snm><fnm>DJT</fnm></au>
    <au><snm>Krause</snm><fnm>J</fnm></au>
  </aug>
  <source>Proc Natl Acad Sci U S A</source>
  <pubdate>2011</pubdate>
  <volume>108</volume>
  <issue>6</issue>
  <fpage>2312</fpage>
  <lpage>-2315</lpage>
  <url>http://dx.doi.org/10.1073/pnas.1007102108</url>
</bibl>

<bibl id="B6">
  <title><p>Collective memory and spatial sorting in animal groups.</p></title>
  <aug>
    <au><snm>Couzin</snm><fnm>ID</fnm></au>
    <au><snm>Krause</snm><fnm>J</fnm></au>
    <au><snm>James</snm><fnm>R</fnm></au>
    <au><snm>Ruxton</snm><fnm>GD</fnm></au>
    <au><snm>Franks</snm><fnm>NR</fnm></au>
  </aug>
  <source>J Theor Biol</source>
  <pubdate>2002</pubdate>
  <volume>218</volume>
  <issue>1</issue>
  <fpage>1</fpage>
  <lpage>-11</lpage>
</bibl>

<bibl id="B7">
  <title><p>The modelling cycle for collective animal behaviour</p></title>
  <aug>
    <au><snm>Sumpter</snm><fnm>DJT</fnm></au>
    <au><snm>Mann</snm><fnm>RP</fnm></au>
    <au><snm>Perna</snm><fnm>A</fnm></au>
  </aug>
  <source>Interface Focus</source>
  <pubdate>2012</pubdate>
  <url>http://rsfs.royalsocietypublishing.org/content/early/2012/08/09/rsfs.2012.0031.abstract</url>
</bibl>

<bibl id="B8">
  <title><p>Interaction Ruling Animal Collective Behaviour Depends on
  Topological rather than Metric Distance: Evidence from a Field
  Study</p></title>
  <aug>
    <au><snm>Ballerini</snm><fnm>M.</fnm></au>
    <au><snm>Cabibbo</snm><fnm>N.</fnm></au>
    <au><snm>Candelier</snm><fnm>R.</fnm></au>
    <au><snm>Cavagna</snm><fnm>A.</fnm></au>
    <au><snm>Cisbani</snm><fnm>E.</fnm></au>
    <au><snm>Giardina</snm><fnm>I.</fnm></au>
    <au><snm>Lecomte</snm><fnm>V.</fnm></au>
    <au><snm>Orlandi</snm><fnm>A.</fnm></au>
    <au><snm>Parisi</snm><fnm>G.</fnm></au>
    <au><snm>Procaccini</snm><fnm>A.</fnm></au>
    <au><snm>Viale</snm><fnm>M.</fnm></au>
    <au><snm>Zdravkovic</snm><fnm>V.</fnm></au>
  </aug>
  <source>PNAS</source>
  <pubdate>2008</pubdate>
  <volume>105</volume>
  <issue>4</issue>
  <fpage>1232</fpage>
  <lpage>-1237</lpage>
</bibl>

<bibl id="B9">
  <title><p>{Inferring individual rules from collective behavior}</p></title>
  <aug>
    <au><snm>Lukeman</snm><fnm>R</fnm></au>
    <au><snm>Li</snm><fnm>YX</fnm></au>
    <au><snm>Edelstein Keshet</snm><fnm>L</fnm></au>
  </aug>
  <source>Proceedings of the National Academy of Sciences</source>
  <pubdate>2010</pubdate>
  <volume>107</volume>
  <issue>28</issue>
  <fpage>12576</fpage>
  <lpage>12580</lpage>
  <url>http://www.pnas.org/content/107/28/12576.abstract</url>
</bibl>

<bibl id="B10">
  <title><p>Using field data to test locust migratory band collective movement
  models</p></title>
  <aug>
    <au><snm>Buhl</snm><fnm>J.</fnm></au>
    <au><snm>Sword</snm><fnm>GA</fnm></au>
    <au><snm>Simpson</snm><fnm>SJ</fnm></au>
  </aug>
  <source>Interface Focus</source>
  <pubdate>2012</pubdate>
  <volume>2</volume>
  <issue>6</issue>
  <fpage>757</fpage>
  <lpage>763</lpage>
  <url>http://rsfs.royalsocietypublishing.org/content/2/6/757.abstract</url>
</bibl>

<bibl id="B11">
  <title><p>Collective Motion due to Individual Escape and Pursuit
  Response</p></title>
  <aug>
    <au><snm>Romanczuk</snm><fnm>P</fnm></au>
    <au><snm>Couzin</snm><fnm>ID</fnm></au>
    <au><snm>Schimansky Geier</snm><fnm>L</fnm></au>
  </aug>
  <source>Phys. Rev. Lett.</source>
  <publisher>American Physical Society</publisher>
  <pubdate>2009</pubdate>
  <volume>102</volume>
  <fpage>010602</fpage>
  <url>http://link.aps.org/doi/10.1103/PhysRevLett.102.010602</url>
</bibl>

<bibl id="B12">
  <title><p>Emergence of Oblong School Shape: Models and Empirical Data of
  Fish</p></title>
  <aug>
    <au><snm>Hemelrijk</snm><fnm>CK</fnm></au>
    <au><snm>Hildenbrandt</snm><fnm>H</fnm></au>
    <au><snm>Reinders</snm><fnm>J</fnm></au>
    <au><snm>Stamhuis</snm><fnm>EJ</fnm></au>
  </aug>
  <source>Ethology</source>
  <publisher>Blackwell Publishing Ltd</publisher>
  <pubdate>2010</pubdate>
  <volume>116</volume>
  <issue>11</issue>
  <fpage>1099</fpage>
  <lpage>-1112</lpage>
  <url>http://dx.doi.org/10.1111/j.1439-0310.2010.01818.x</url>
</bibl>

<bibl id="B13">
  <title><p>Inferring the structure and dynamics of interactions in schooling
  fish.</p></title>
  <aug>
    <au><snm>Katz</snm><fnm>Y</fnm></au>
    <au><snm>Tunstrøm</snm><fnm>K</fnm></au>
    <au><snm>Ioannou</snm><fnm>CC</fnm></au>
    <au><snm>Huepe</snm><fnm>C</fnm></au>
    <au><snm>Couzin</snm><fnm>ID</fnm></au>
  </aug>
  <source>Proc Natl Acad Sci U S A</source>
  <pubdate>2011</pubdate>
  <volume>108</volume>
  <issue>46</issue>
  <fpage>18720</fpage>
  <lpage>-18725</lpage>
  <url>http://dx.doi.org/10.1073/pnas.1107583108</url>
</bibl>

<bibl id="B14">
  <title><p>Inferring the rules of interaction of shoaling fish.</p></title>
  <aug>
    <au><snm>Herbert Read</snm><fnm>JE</fnm></au>
    <au><snm>Perna</snm><fnm>A</fnm></au>
    <au><snm>Mann</snm><fnm>RP</fnm></au>
    <au><snm>Schaerf</snm><fnm>TM</fnm></au>
    <au><snm>Sumpter</snm><fnm>DJT</fnm></au>
    <au><snm>Ward</snm><fnm>AJW</fnm></au>
  </aug>
  <source>Proc Natl Acad Sci U S A</source>
  <pubdate>2011</pubdate>
  <volume>108</volume>
  <issue>46</issue>
  <fpage>18726</fpage>
  <lpage>-18731</lpage>
  <url>http://dx.doi.org/10.1073/pnas.1109355108</url>
</bibl>

<bibl id="B15">
  <title><p>Interaction rules underlying group decisions in homing
  pigeons</p></title>
  <aug>
    <au><snm>Pettit</snm><fnm>B</fnm></au>
    <au><snm>Perna</snm><fnm>A</fnm></au>
    <au><snm>Biro</snm><fnm>D</fnm></au>
    <au><snm>Sumpter</snm><fnm>DJT</fnm></au>
  </aug>
  <source>Journal of The Royal Society Interface</source>
  <pubdate>2013</pubdate>
  <volume>10</volume>
  <issue>89</issue>
  <url>http://rsif.royalsocietypublishing.org/content/10/89/20130529.abstract</url>
</bibl>

<bibl id="B16">
  <title><p>Hierarchical group dynamics in pigeon flocks.</p></title>
  <aug>
    <au><snm>Nagy</snm><fnm>M</fnm></au>
    <au><snm>Akos</snm><fnm>Z</fnm></au>
    <au><snm>Biro</snm><fnm>D</fnm></au>
    <au><snm>Vicsek</snm><fnm>T</fnm></au>
  </aug>
  <source>Nature</source>
  <pubdate>2010</pubdate>
  <volume>464</volume>
  <issue>7290</issue>
  <fpage>890</fpage>
  <lpage>-893</lpage>
</bibl>

<bibl id="B17">
  <title><p>Moving and staying together without a leader</p></title>
  <aug>
    <au><snm>Gregoire</snm><fnm>G</fnm></au>
    <au><snm>Chate</snm><fnm>H</fnm></au>
    <au><snm>Tu</snm><fnm>Y</fnm></au>
  </aug>
  <source>Physica D: Nonlinear Phenomena</source>
  <pubdate>2003</pubdate>
  <volume>181</volume>
  <issue>3--4</issue>
  <fpage>157</fpage>
  <lpage>-170</lpage>
  <url>http://www.sciencedirect.com/science/article/pii/S0167278903001027</url>
</bibl>

<bibl id="B18">
  <title><p>The three-dimensional structure of fish schools</p></title>
  <aug>
    <au><snm>Partridge</snm><fnm>BL</fnm></au>
    <au><snm>Pitcher</snm><fnm>T</fnm></au>
    <au><snm>Cullen</snm><fnm>JM</fnm></au>
    <au><snm>Wilson</snm><fnm>J</fnm></au>
  </aug>
  <source>Behavioral Ecology and Sociobiology</source>
  <pubdate>1980</pubdate>
  <volume>6</volume>
  <issue>4</issue>
  <fpage>277</fpage>
  <lpage>-288</lpage>
  <url>http://dx.doi.org/10.1007/BF00292770</url>
  <note>10.1007/BF00292770</note>
</bibl>

<bibl id="B19">
  <title><p>Collective motion from local attraction</p></title>
  <aug>
    <au><snm>Strombom</snm><fnm>D</fnm></au>
  </aug>
  <source>Journal of Theoretical Biology</source>
  <publisher>24-28 OVAL RD, LONDON NW1 7DX, ENGLAND: ACADEMIC PRESS LTD-
  ELSEVIER SCIENCE LTD</publisher>
  <pubdate>2011</pubdate>
  <volume>283</volume>
  <issue>1</issue>
  <fpage>145</fpage>
  <lpage>-151</lpage>
</bibl>

<bibl id="B20">
  <title><p>Controlling active self-assembly through broken particle-shape
  symmetry</p></title>
  <aug>
    <au><snm>Wensink</snm><fnm>H. H.</fnm></au>
    <au><snm>Kantsler</snm><fnm>V.</fnm></au>
    <au><snm>Goldstein</snm><fnm>R. E.</fnm></au>
    <au><snm>Dunkel</snm><fnm>J.</fnm></au>
  </aug>
  <source>Phys. Rev. E</source>
  <publisher>American Physical Society</publisher>
  <pubdate>2014</pubdate>
  <volume>89</volume>
  <fpage>010302</fpage>
  <url>http://link.aps.org/doi/10.1103/PhysRevE.89.010302</url>
</bibl>

<bibl id="B21">
  <title><p>Emergent states in dense systems of active rods: from swarming to
  turbulence</p></title>
  <aug>
    <au><snm>Wensink</snm><fnm>H H</fnm></au>
    <au><snm>Löwen</snm><fnm>H</fnm></au>
  </aug>
  <source>Journal of Physics: Condensed Matter</source>
  <pubdate>2012</pubdate>
  <volume>24</volume>
  <issue>46</issue>
  <fpage>464130</fpage>
  <url>http://stacks.iop.org/0953-8984/24/i=46/a=464130</url>
</bibl>

<bibl id="B22">
  <title><p>The increased efficiency of fish swimming in a school</p></title>
  <aug>
    <au><snm>Hemelrijk</snm><fnm>CK</fnm></au>
    <au><snm>Reid</snm><fnm>DAP</fnm></au>
    <au><snm>Hildenbrandt</snm><fnm>H</fnm></au>
    <au><snm>Padding</snm><fnm>JT</fnm></au>
  </aug>
  <source>Fish and Fisheries</source>
  <pubdate>2014</pubdate>
  <fpage>n/a</fpage>
  <lpage>-n/a</lpage>
  <url>http://dx.doi.org/10.1111/faf.12072</url>
</bibl>

<bibl id="B23">
  <title><p>Objectively identifying landmark use and predicting flight
  trajectories of the homing pigeon using Gaussian processes</p></title>
  <aug>
    <au><snm>Mann</snm><fnm>R</fnm></au>
    <au><snm>Freeman</snm><fnm>R</fnm></au>
    <au><snm>Osborne</snm><fnm>M</fnm></au>
    <au><snm>Garnett</snm><fnm>R</fnm></au>
    <au><snm>Armstrong</snm><fnm>C</fnm></au>
    <au><snm>Meade</snm><fnm>J</fnm></au>
    <au><snm>Biro</snm><fnm>D</fnm></au>
    <au><snm>Guilford</snm><fnm>T</fnm></au>
    <au><snm>Roberts</snm><fnm>S</fnm></au>
  </aug>
  <source>Journal of The Royal Society Interface</source>
  <pubdate>2011</pubdate>
  <volume>8</volume>
  <issue>55</issue>
  <fpage>210</fpage>
  <lpage>219</lpage>
  <url>http://rsif.royalsocietypublishing.org/content/8/55/210.abstract</url>
</bibl>

<bibl id="B24">
  <title><p>Group decisions and individual differences: route fidelity predicts
  flight leadership in homing pigeons (Columba livia)</p></title>
  <aug>
    <au><snm>Freeman</snm><fnm>R</fnm></au>
    <au><snm>Mann</snm><fnm>R</fnm></au>
    <au><snm>Guilford</snm><fnm>T</fnm></au>
    <au><snm>Biro</snm><fnm>D</fnm></au>
  </aug>
  <source>Biology Letters</source>
  <pubdate>2011</pubdate>
  <volume>7</volume>
  <issue>1</issue>
  <fpage>63</fpage>
  <lpage>66</lpage>
  <url>http://rsbl.royalsocietypublishing.org/content/7/1/63.abstract</url>
</bibl>

<bibl id="B25">
  <title><p>Distinguishing social from nonsocial navigation in moving animal
  groups</p></title>
  <aug>
    <au><snm>Bode</snm><fnm>NW</fnm></au>
    <au><snm>Franks</snm><fnm>DW</fnm></au>
    <au><snm>Wood</snm><fnm>AJ</fnm></au>
    <au><snm>Piercy</snm><fnm>JJ</fnm></au>
    <au><snm>Croft</snm><fnm>DP</fnm></au>
    <au><snm>Codling</snm><fnm>EA</fnm></au>
  </aug>
  <source>The American Naturalist</source>
  <publisher>JSTOR</publisher>
  <pubdate>2012</pubdate>
  <volume>179</volume>
  <issue>5</issue>
  <fpage>621</fpage>
  <lpage>-632</lpage>
</bibl>

<bibl id="B26">
  <title><p>Follow me! I’m a leader if you do; I’m a failed initiator if
  you don’t?</p></title>
  <aug>
    <au><snm>King</snm><fnm>AJ</fnm></au>
  </aug>
  <source>Behavioural Processes</source>
  <pubdate>2010</pubdate>
  <volume>84</volume>
  <issue>3</issue>
  <fpage>671</fpage>
  <lpage>674</lpage>
  <url>http://www.sciencedirect.com/science/article/pii/S0376635710000975</url>
  <note>Special section: Collective movements Decision-making processes within
  groups</note>
</bibl>

<bibl id="B27">
  <title><p>Leadership in fish shoals</p></title>
  <aug>
    <au><snm>Krause</snm><fnm>J</fnm></au>
    <au><snm>Hoare</snm><fnm>D</fnm></au>
    <au><snm>Krause</snm><fnm>S</fnm></au>
    <au><snm>Hemelrijk</snm><fnm>CK</fnm></au>
    <au><snm>Rubenstein</snm><fnm>DI</fnm></au>
  </aug>
  <source>Fish and Fisheries</source>
  <publisher>Blackwell Science Ltd</publisher>
  <pubdate>2000</pubdate>
  <volume>1</volume>
  <issue>1</issue>
  <fpage>82</fpage>
  <lpage>-89</lpage>
</bibl>

<bibl id="B28">
  <title><p>Aerobic capacity influences the spatial position of individuals
  within fish schools</p></title>
  <aug>
    <au><snm>Killen</snm><fnm>SS</fnm></au>
    <au><snm>Marras</snm><fnm>S</fnm></au>
    <au><snm>Steffensen</snm><fnm>JF</fnm></au>
    <au><snm>McKenzie</snm><fnm>DJ</fnm></au>
  </aug>
  <source>Proceedings of the Royal Society B: Biological Sciences</source>
  <pubdate>2012</pubdate>
  <volume>279</volume>
  <issue>1727</issue>
  <fpage>357</fpage>
  <lpage>364</lpage>
  <url>http://rspb.royalsocietypublishing.org/content/279/1727/357.abstract</url>
</bibl>

</refgrp>
} % end of \BMCxmlcomment

%%%%%%%%%%%%%%%%%%%%%%%%%%%%%%%%%%%
%%                               %%
%% Figures                       %%
%%                               %%
%% NB: this is for captions and  %%
%% Titles. All graphics must be  %%
%% submitted separately and NOT  %%
%% included in the Tex document  %%
%%                               %%
%%%%%%%%%%%%%%%%%%%%%%%%%%%%%%%%%%%

%%
%% Do not use \listoffigures as most will be included as separate files

\section*{Figures}
%  \begin{figure}[h!]
%  \caption{\csentence{Sample figure title.}
%      A short description of the figure content
%      should go here.}
%      \end{figure}
%
%\begin{figure}[h!]
%  \caption{\csentence{Sample figure title.}
%      Figure legend text.}
%      \end{figure}
\begin{figure}[!tbh]
\begin{center}
\includegraphics[width=0.7\columnwidth]{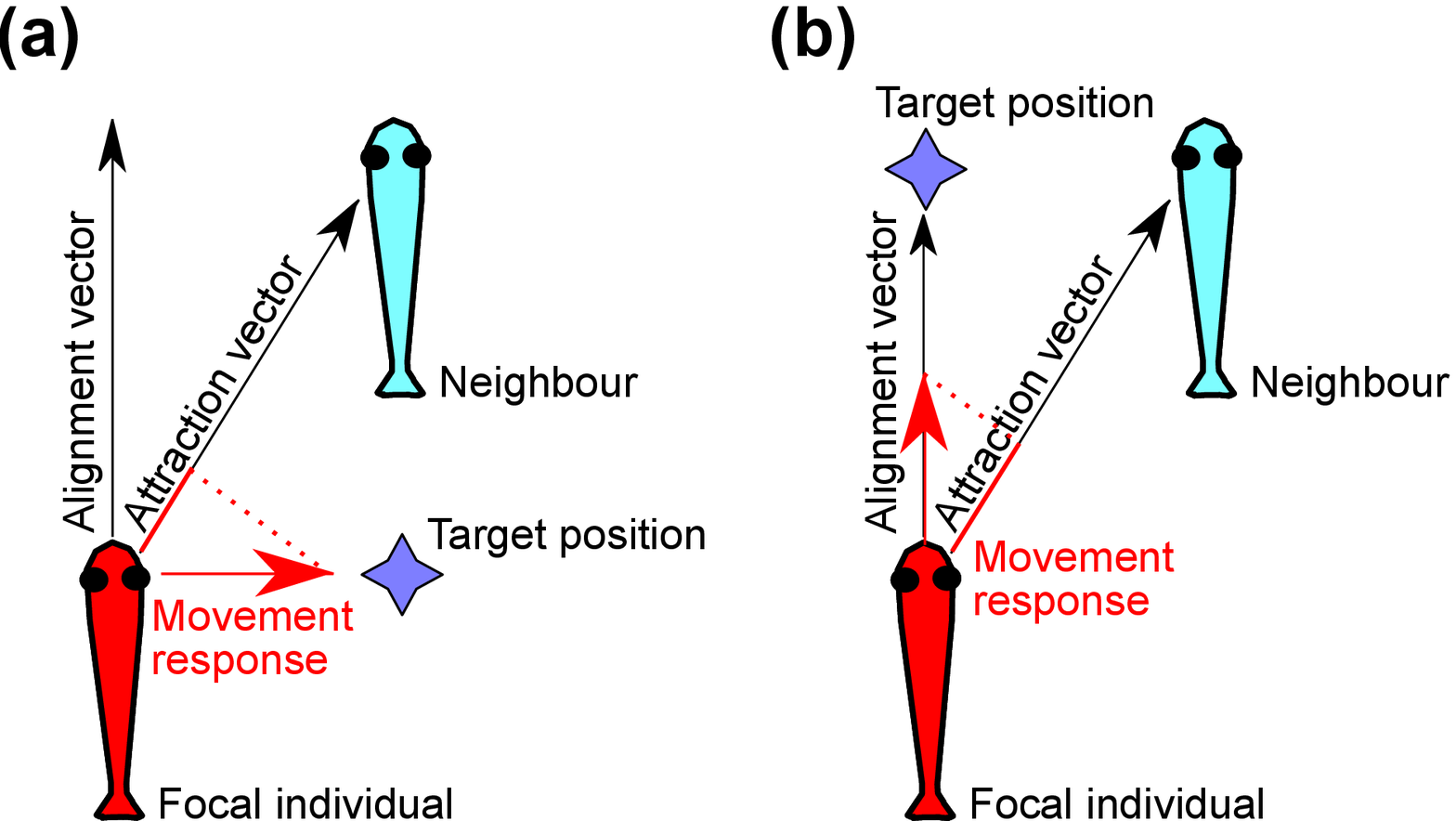}
\end{center}
\caption{
\csentence{Illustration of the interactions.}  The focal fish (in red) aims at keeping a stable target position relative to its neighbour. In {\bf (a)} this target position is behind the neighbour, while in {\bf (b)} it is on the side of the neighbour. The movement in the direction of the target can be interpreted in terms of attraction or repulsion response if it has a projection onto the attraction/repulsion vector pointing in the direction of the neighbour. If the movement response has a component along the direction parallel to the neighbour (the alignment vector), it can also be interpreted as alignment. In general, the attraction / repulsion vector and the alignment vector are not orthogonal to each other, and in the particular case of aligned individuals with target positions in front or behind, the attraction and alignment vectors are not even linearly independent.
}
\label{fig:illustrated_interactions}
\end{figure}

\begin{figure}[!tbh]
\begin{center}
\includegraphics[width=0.7\columnwidth]{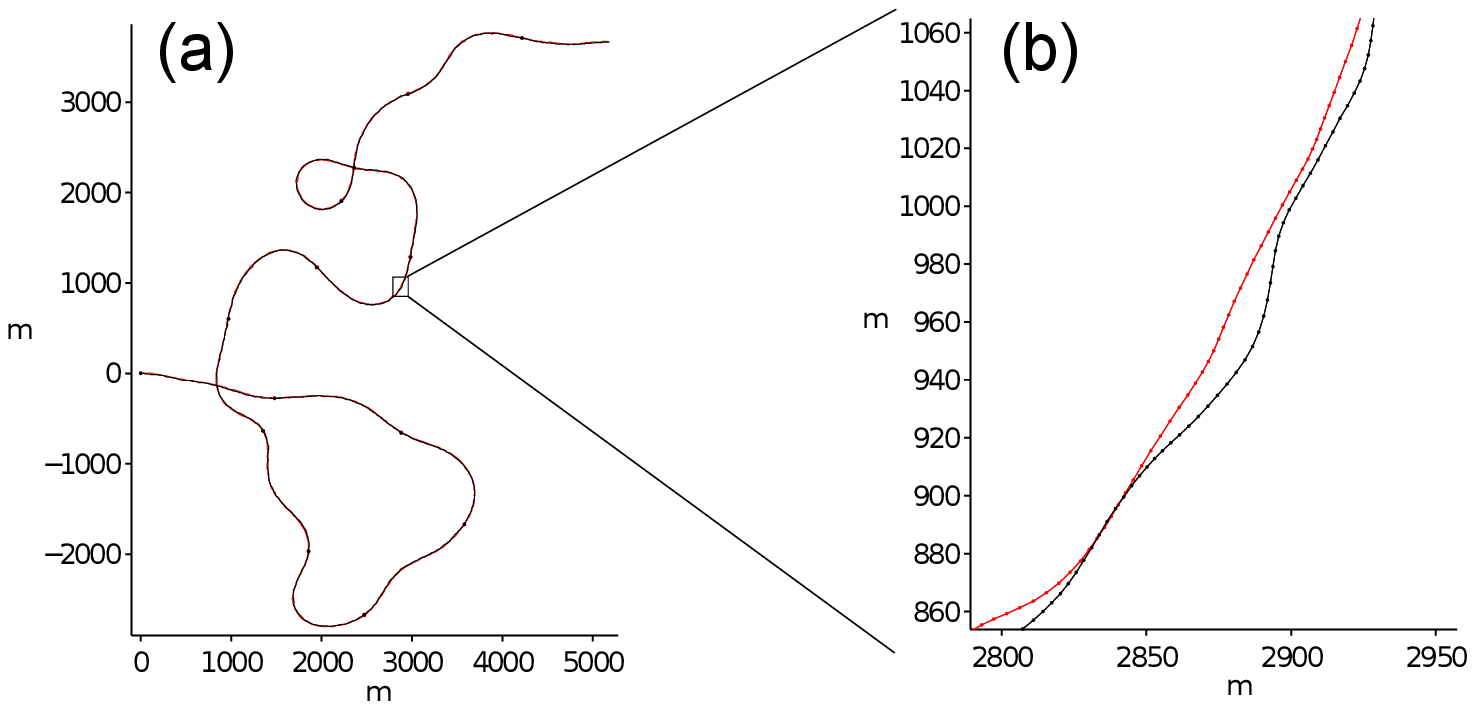}
\end{center}
\caption{
\csentence{Example of generated trajectories for two particles moving side by side.}  {\bf (a)} Complete trajectory of $2^{12}$ steps. The larger dots (visible when zooming in the figure) indicate the scale for temporal correlation $C_T$ (=300 steps) used for generating the trajectories. {\bf (b)} Zoom on a smaller portion of trajectory to illustrate the recorded positions of both individuals. Each dot represents the position at one different time step.
}
\label{fig:trajectory}
\end{figure}

\begin{figure*}[!tb]
\begin{minipage}[t]{1\textwidth}
\begin{tabular}{ccc}
\includegraphics[width=0.3\textwidth]{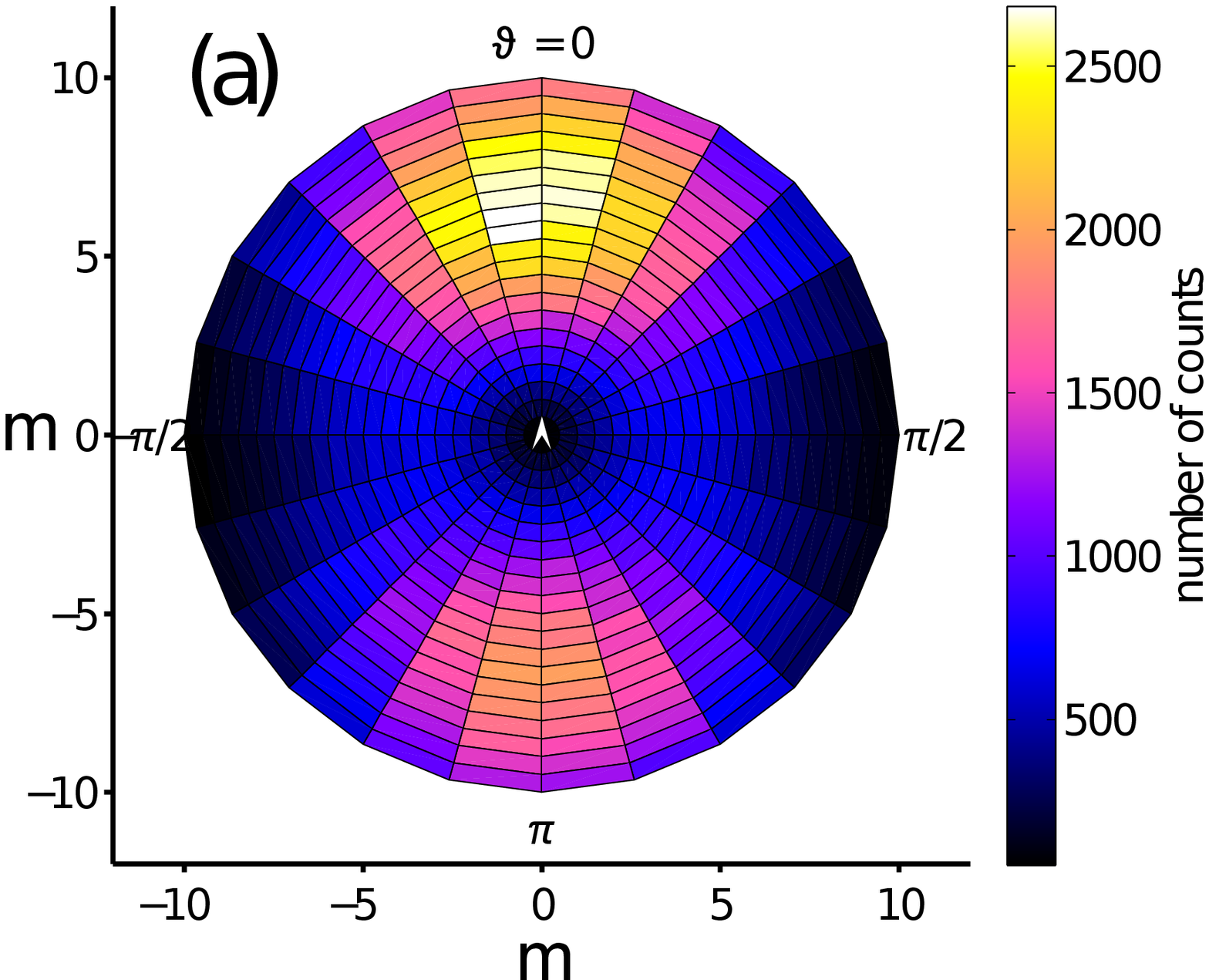} &
\includegraphics[width=0.3\textwidth]{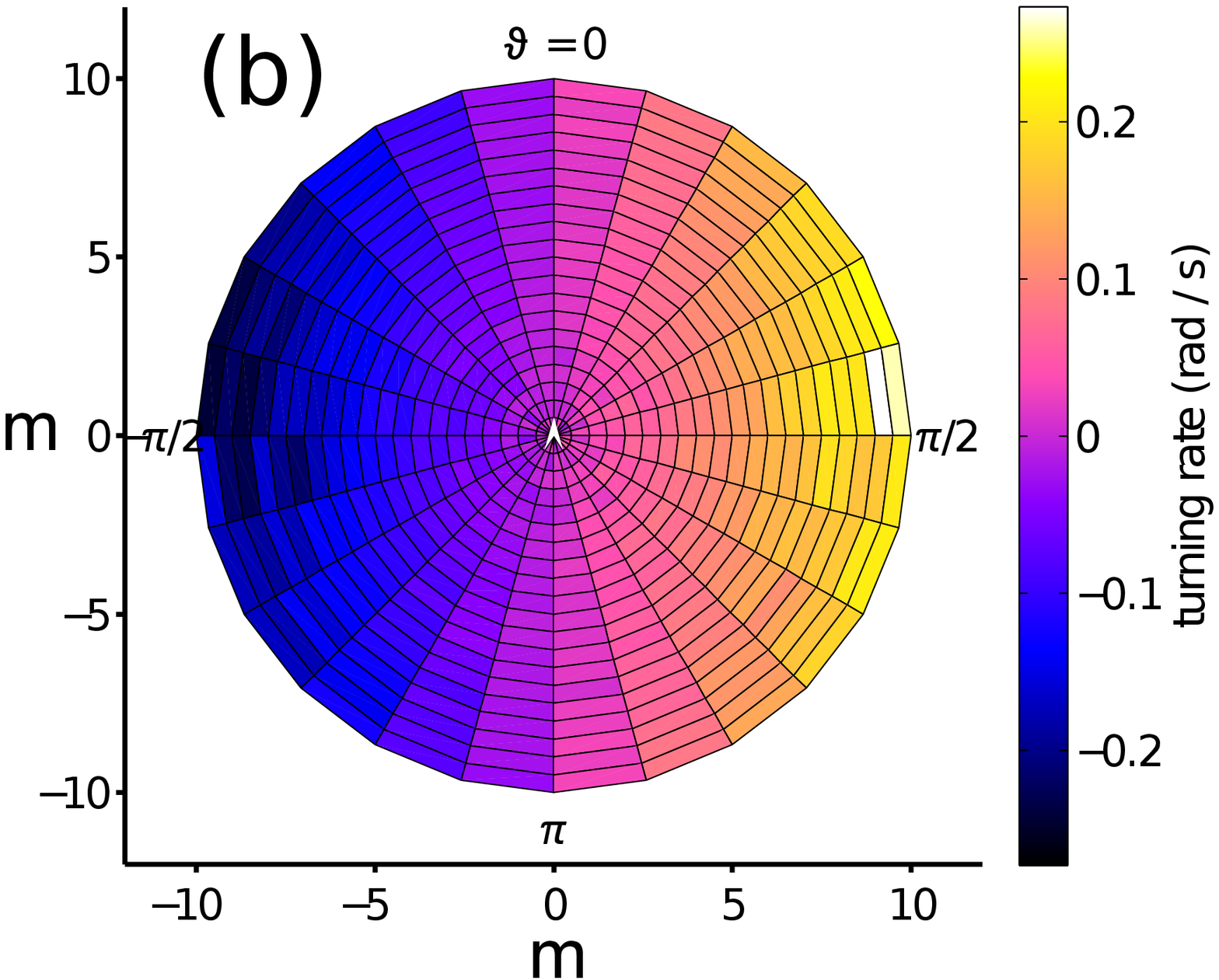} &
\includegraphics[width=0.3\textwidth]{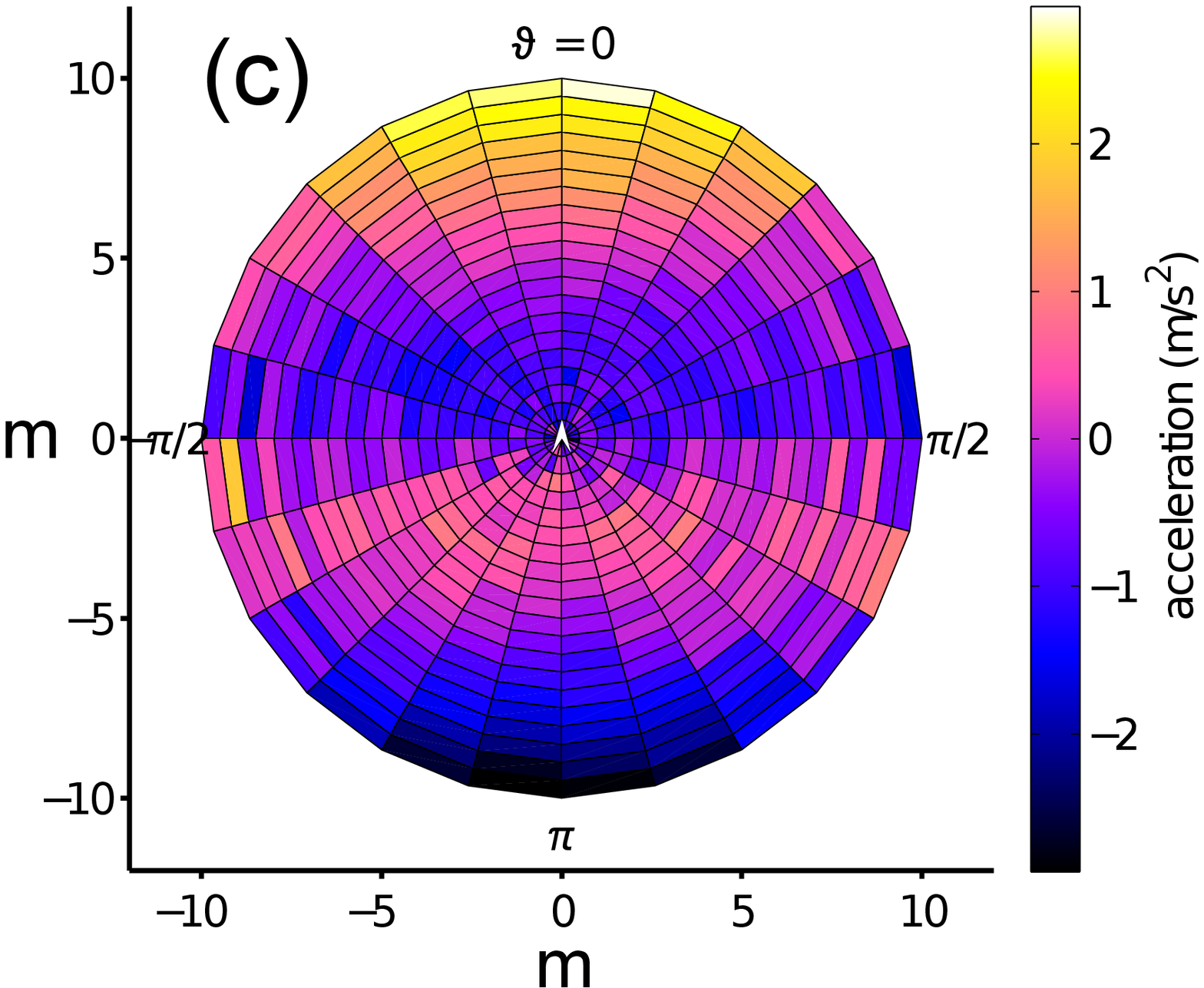} \\
\includegraphics[width=0.3\textwidth]{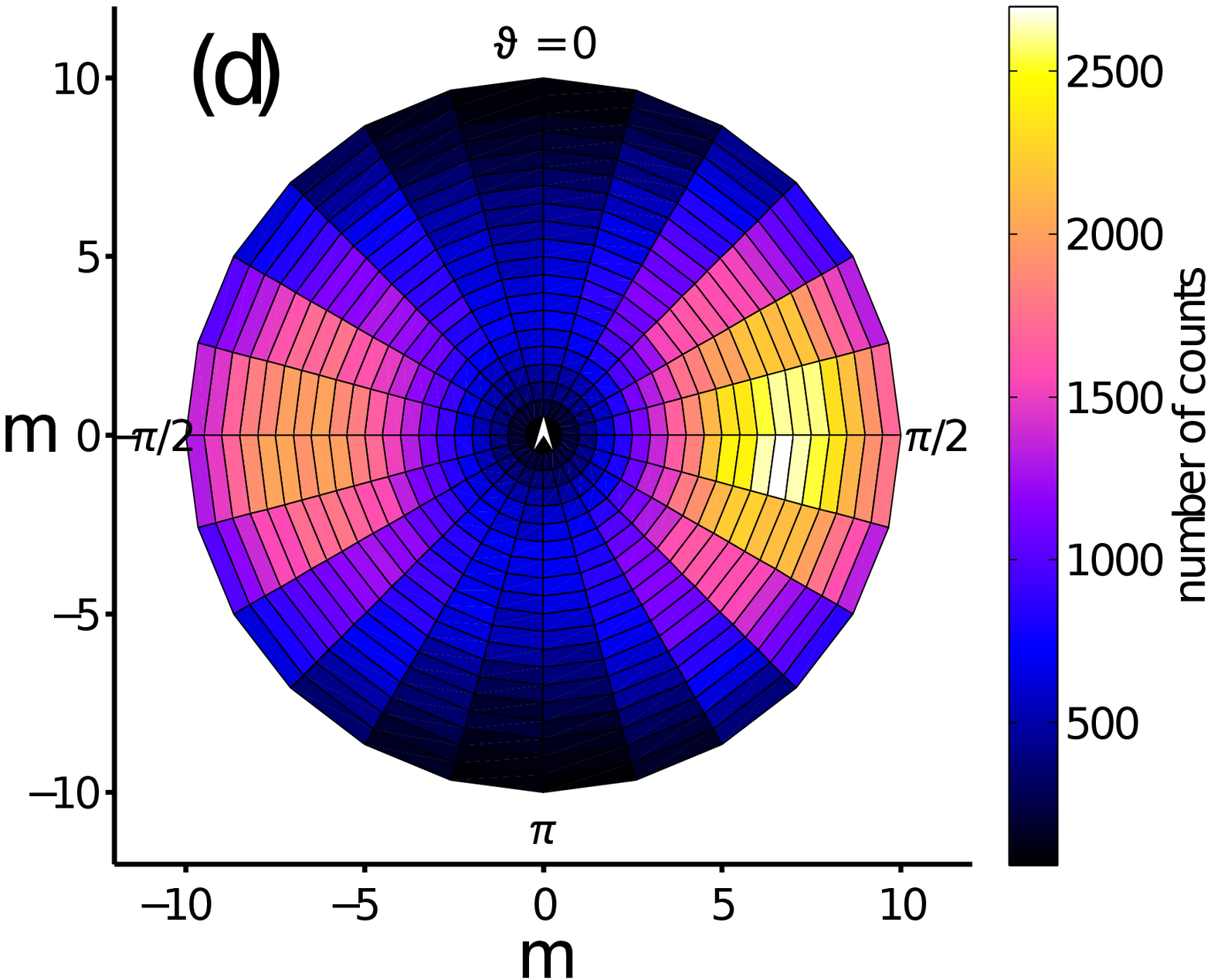} &
\includegraphics[width=0.3\textwidth]{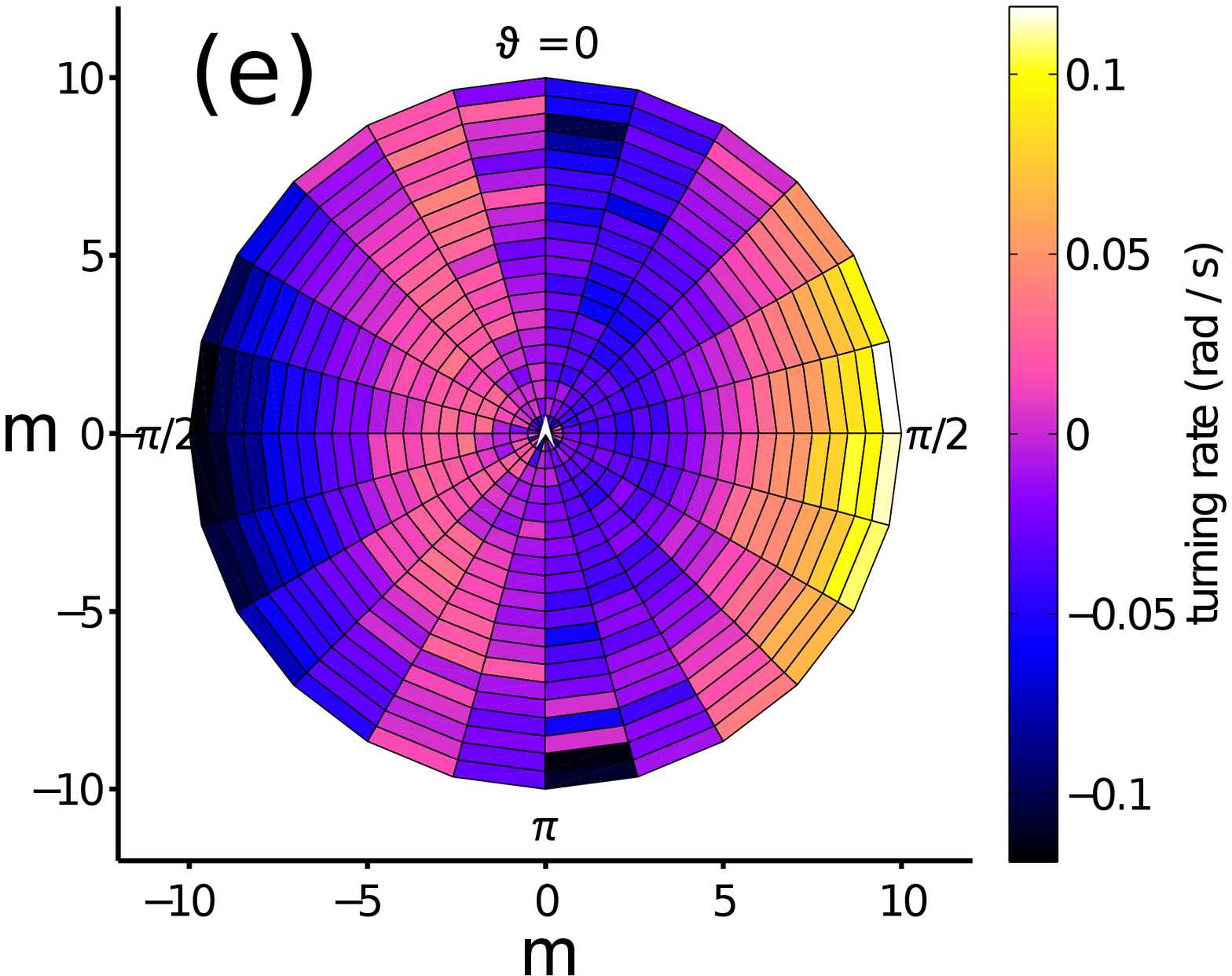} &
\includegraphics[width=0.3\textwidth]{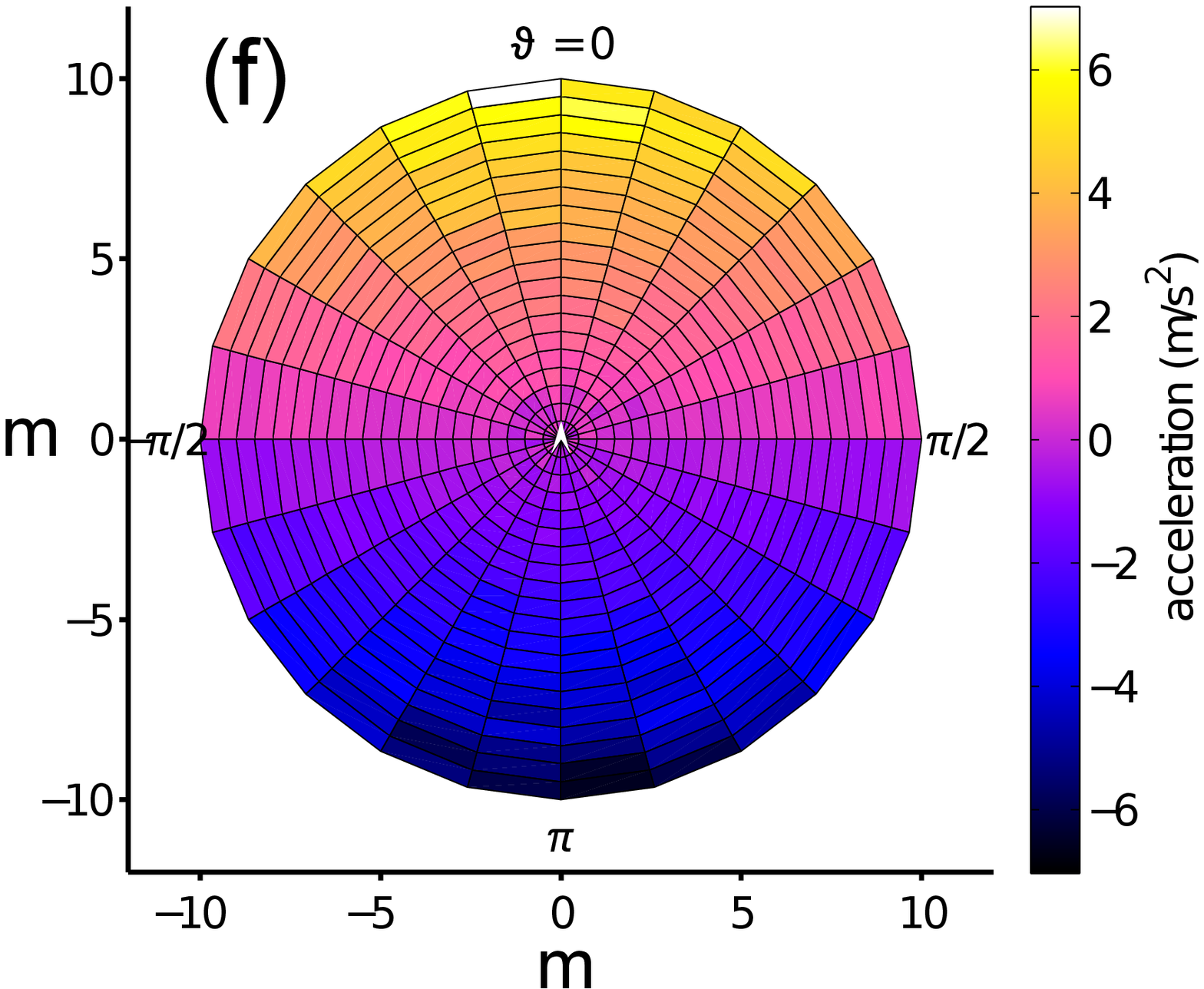} \\
\end{tabular}
\end{minipage}
\caption{
\csentence{Inferred interaction rules as a function of distance and direction to the neighbour}. {\bf Top row} Individuals moving in a front-back configuration. {\bf Bottom row}. Individuals moving side by side. {\bf (a)} and {\bf (d)}  Number of counts of the neighbour within each cell of the polar grid. The positions at which the neighbour is most frequently observed match those imposed when generating the trajectories. {\bf b} Turning response. When the individuals move in a front-back configuration, turning always happens in the direction of the neighbour. {\bf c} Acceleration response for individuals moving in a front-back configuration. Close-by neighbours elicit a repulsive response, with an acceleration of the opposite sign. {\bf e} Turning response of individuals moving side by side. Repulsion is mediated through turning away from the neighbour. {\bf f} Acceleration response. For individuals moving side by side, acceleration is always positive when the neighbour is in front and negative when the neighbour is behind.
}
\label{fig:apparent_rules_of_motion_of_particles_from_trajectory_analysis}
\end{figure*}

\begin{figure}[!tb]
\begin{minipage}[t]{1\textwidth}
\begin{tabular}{cc}
\includegraphics[width=0.45\textwidth]{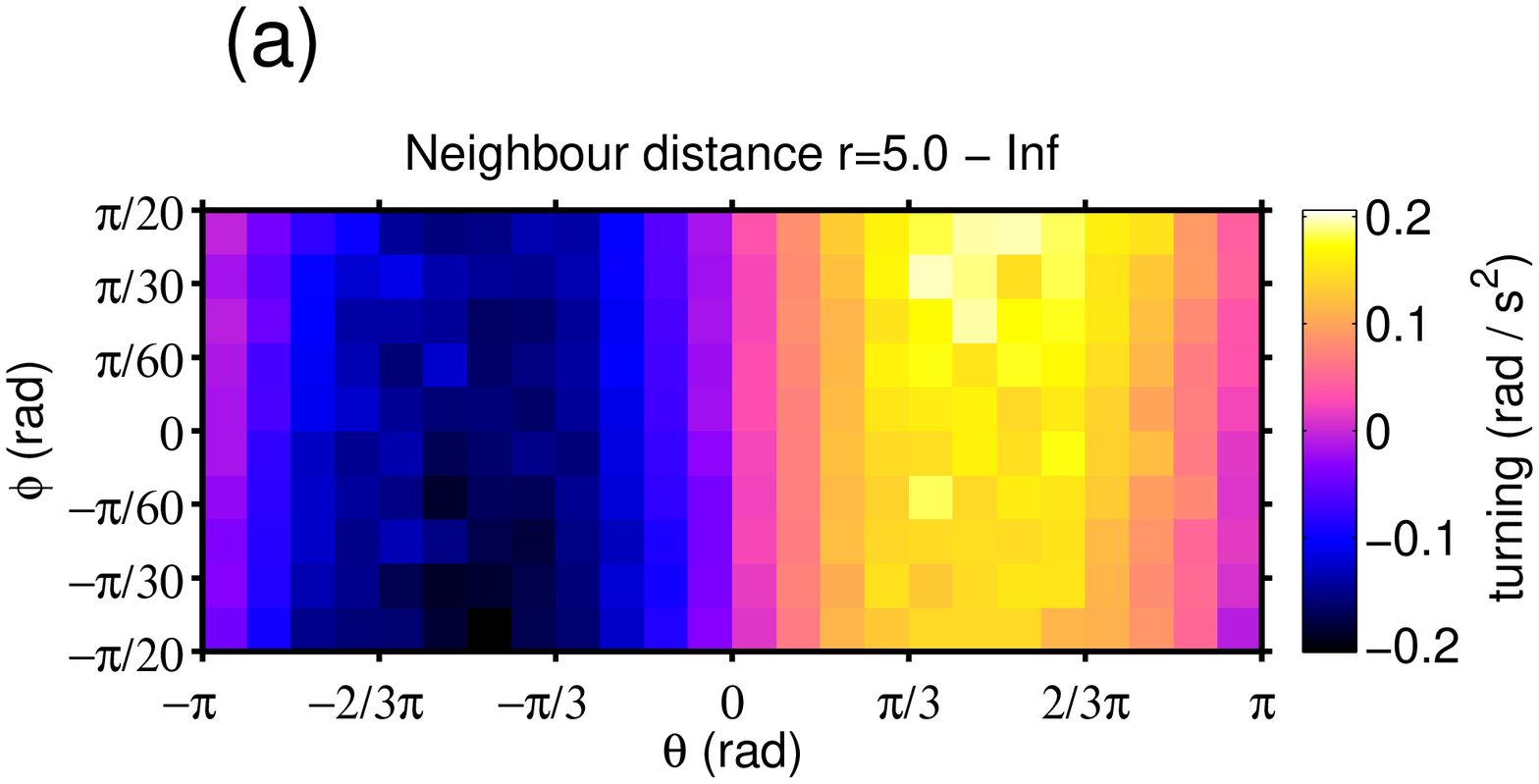} &
\includegraphics[width=0.45\textwidth]{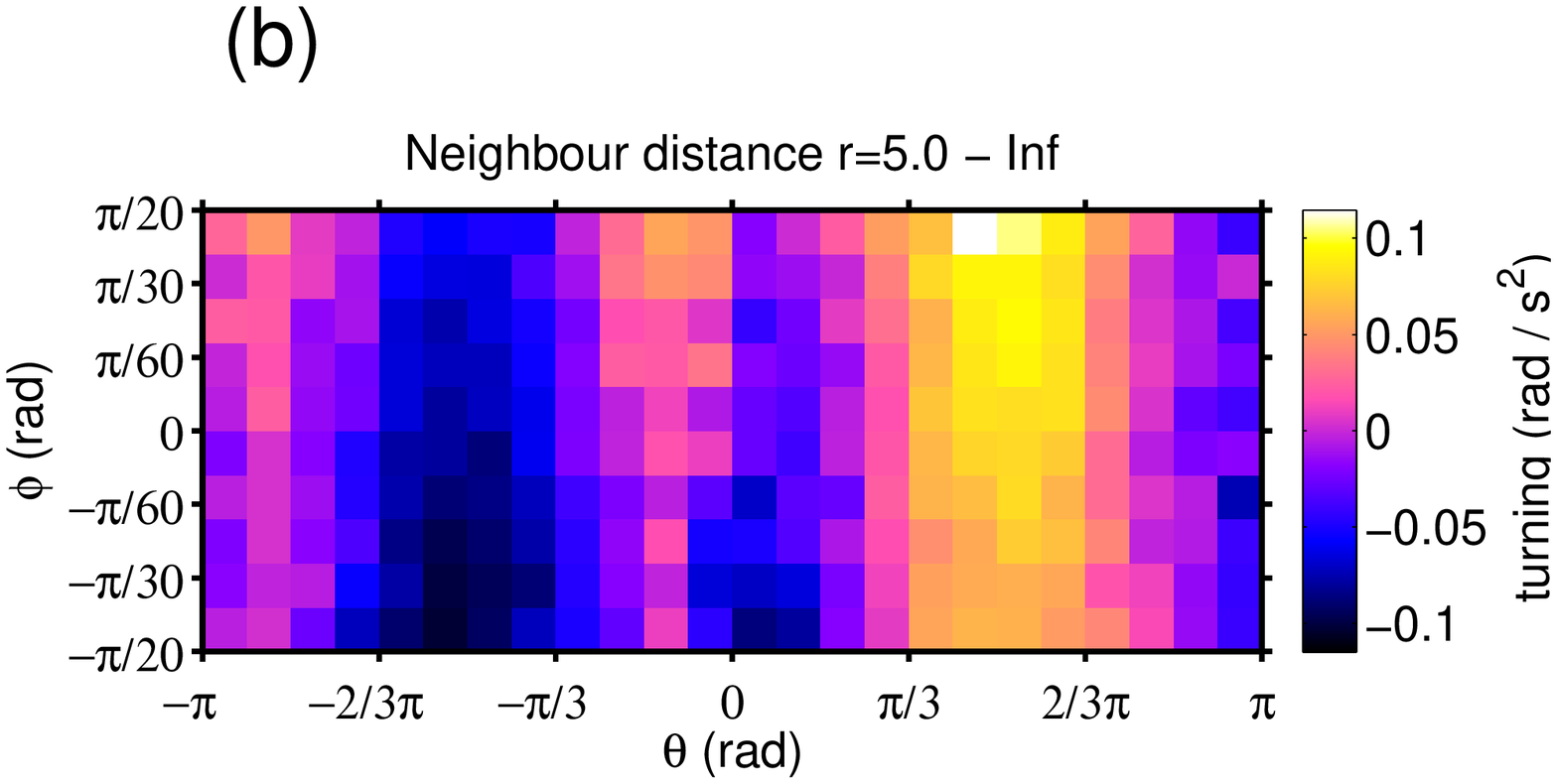} \\
\includegraphics[width=0.45\textwidth]{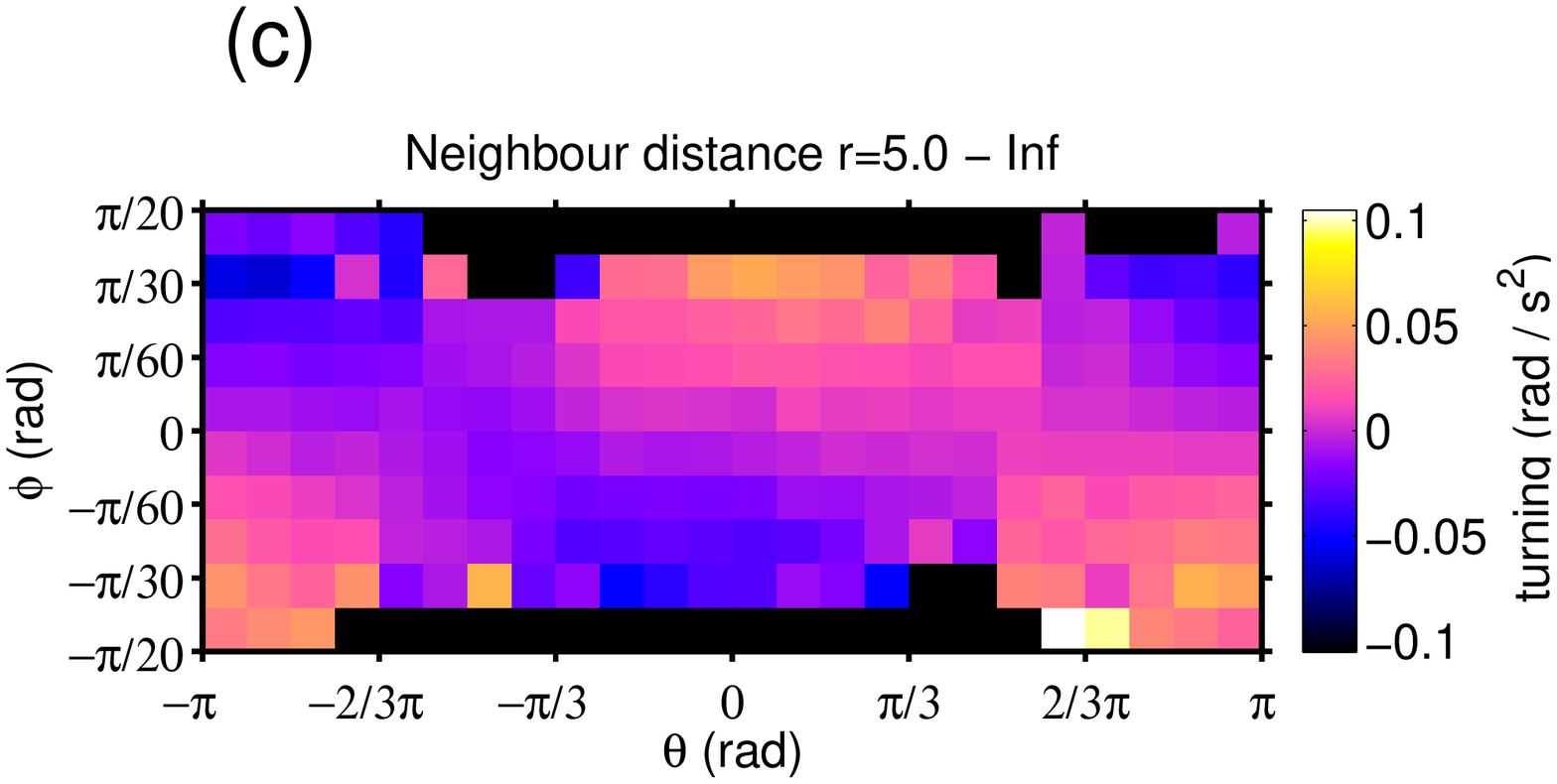} &
\includegraphics[width=0.45\textwidth]{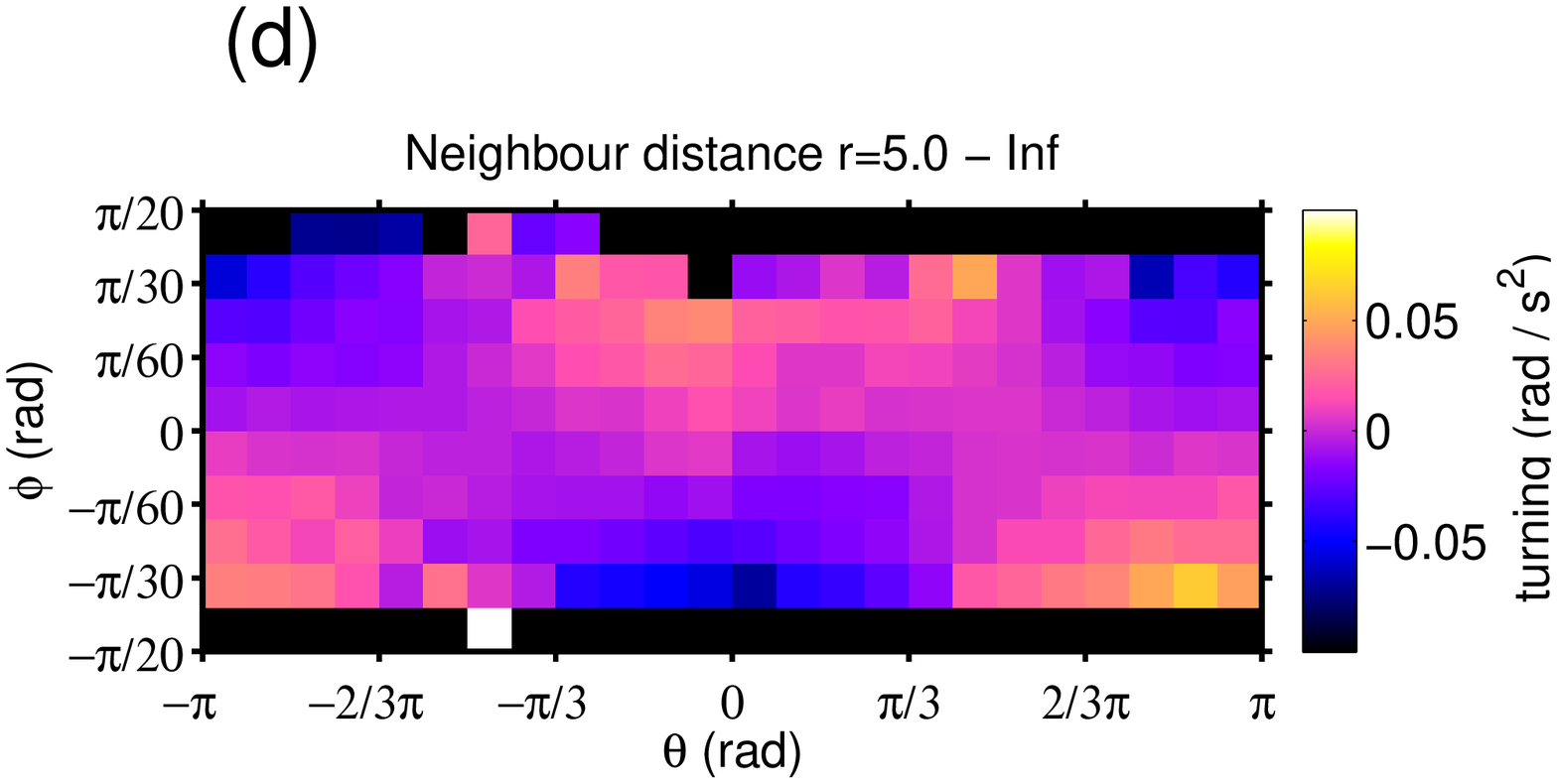}
\end{tabular}
\end{minipage}
\caption{
\csentence{Relative effect of `attraction' and `alignment'.} The figures represent the average turning angle of the focal individual in response to the direction ($\theta$) and relative orientation ($\phi$) of the neighbour, limited to situations in which the neighbour is in the attraction zone (at a distance $r>5\textrm{m}$). {\bf (a)} The two particles fly in a front-back configuration. {\bf (b)} Particles flying side by side; {\bf (c)} Same as (a), but with increased temporal autocorrelation of noise around the target position ($C_D = 100 \textrm{steps}$, while it was $C_D = 20 \textrm{steps}$ in the previous plots). {\bf (d)} Same as (b), with increased temporal autocorrelation of noise.
}
\label{fig:attraction_vs_alignment_of_particles_from_trajectory_analysis}
\end{figure}

\begin{figure}[!tb]
\begin{minipage}[t]{1\columnwidth}
\begin{tabular}{cc}
\includegraphics[width=0.5\columnwidth]{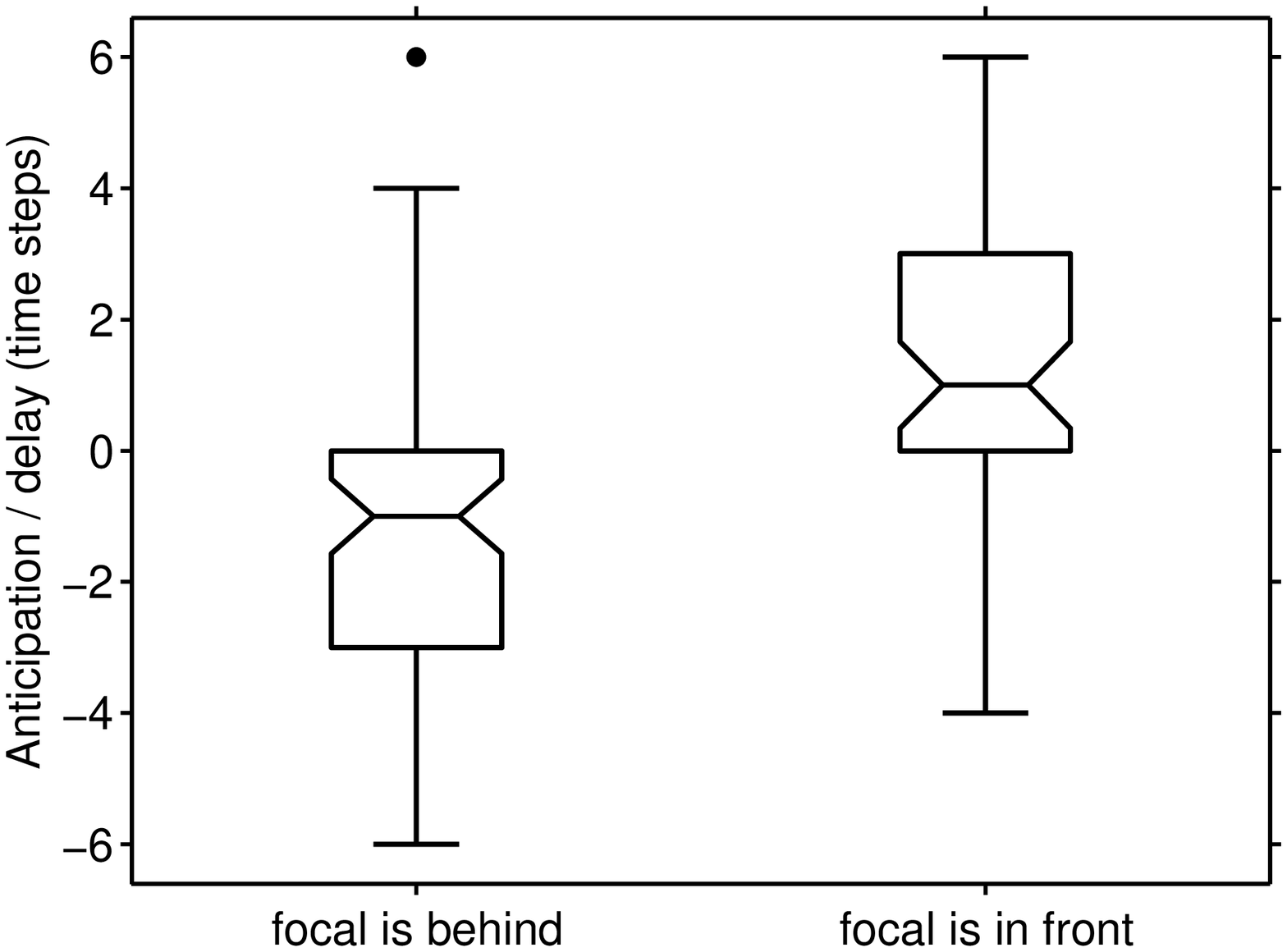} &
\includegraphics[width=0.5\columnwidth]{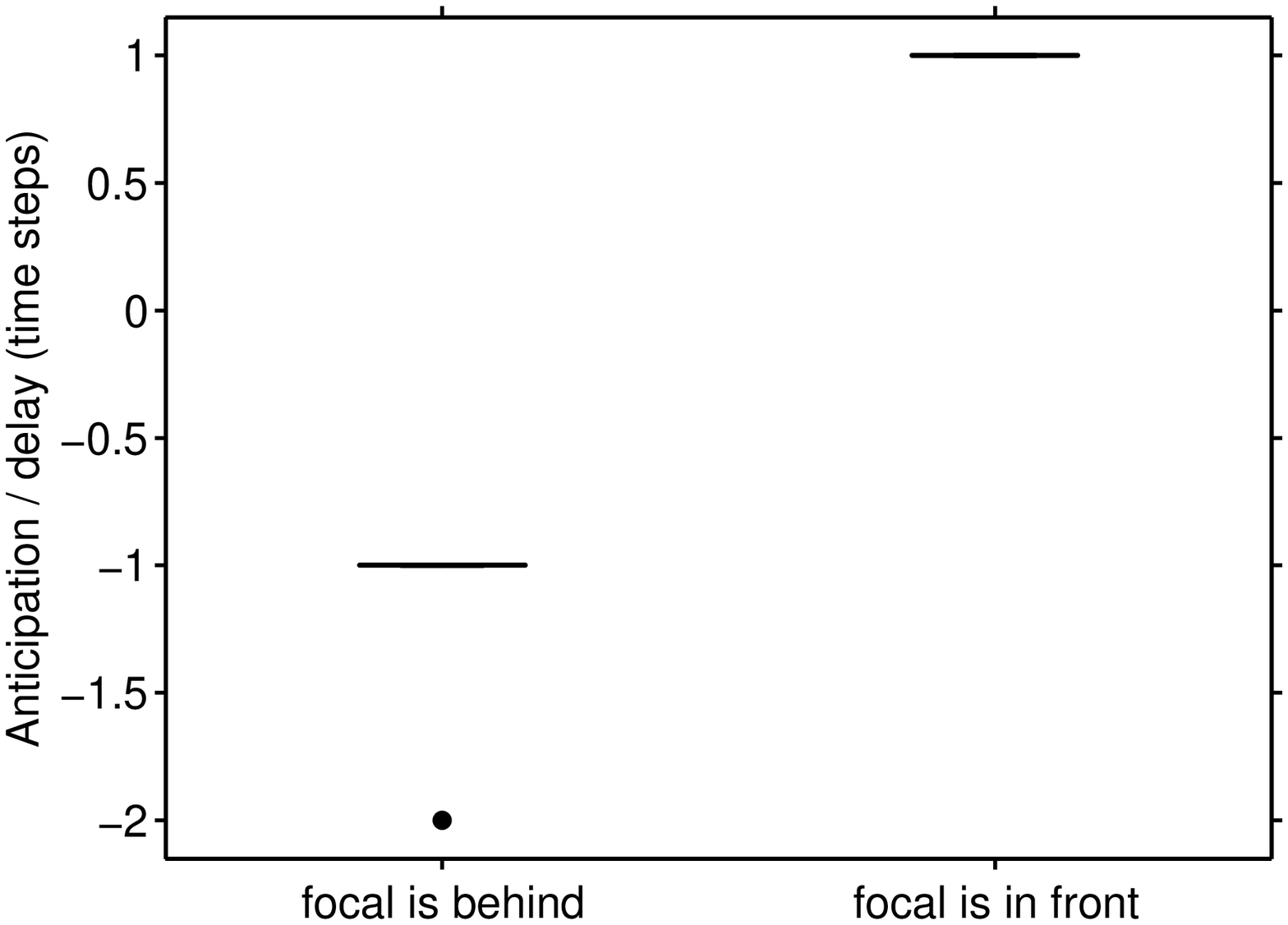}
\end{tabular}
\end{minipage}
\caption{
\csentence{Directional correlation delay vs. position in the group.} Each boxplot represents the distribution of directional correlation delays $\tau^*$ over simulated trajectories. The box on the left indicates trajectories in which the focal individual was in front; while the box on the right indicates those where the focal individual was behind. In our convention, positive values of the correlation delay $\tau^*$ indicate that the focal individual anticipates the changes of direction of its partner. When the individuals fly in a front-back configuration, measures of directional correlation indicate that the individual in front anticipates the turns of its neighbour. {\bf Left} Individuals flying in a front-back configuration, temporal autocorrelation of the noise is short ($C_D = 20 \textrm{steps}$); 120 simulated trajectories {\bf Right} Same simulation parameters as for the figure on the left, but with longer temporal autocorrelation of noise ($C_D = 100 \textrm{steps}$). Note that in this case the variability is extremely reduced and $\tau^*$ was equal to $\pm 1$ in all but one simulation.
}
\label{fig:directional_correlation_delay}
\end{figure}

\begin{figure}[!tb]
\begin{minipage}[t]{1\textwidth}
\begin{tabular}{cc}
\includegraphics[width=0.45\textwidth]{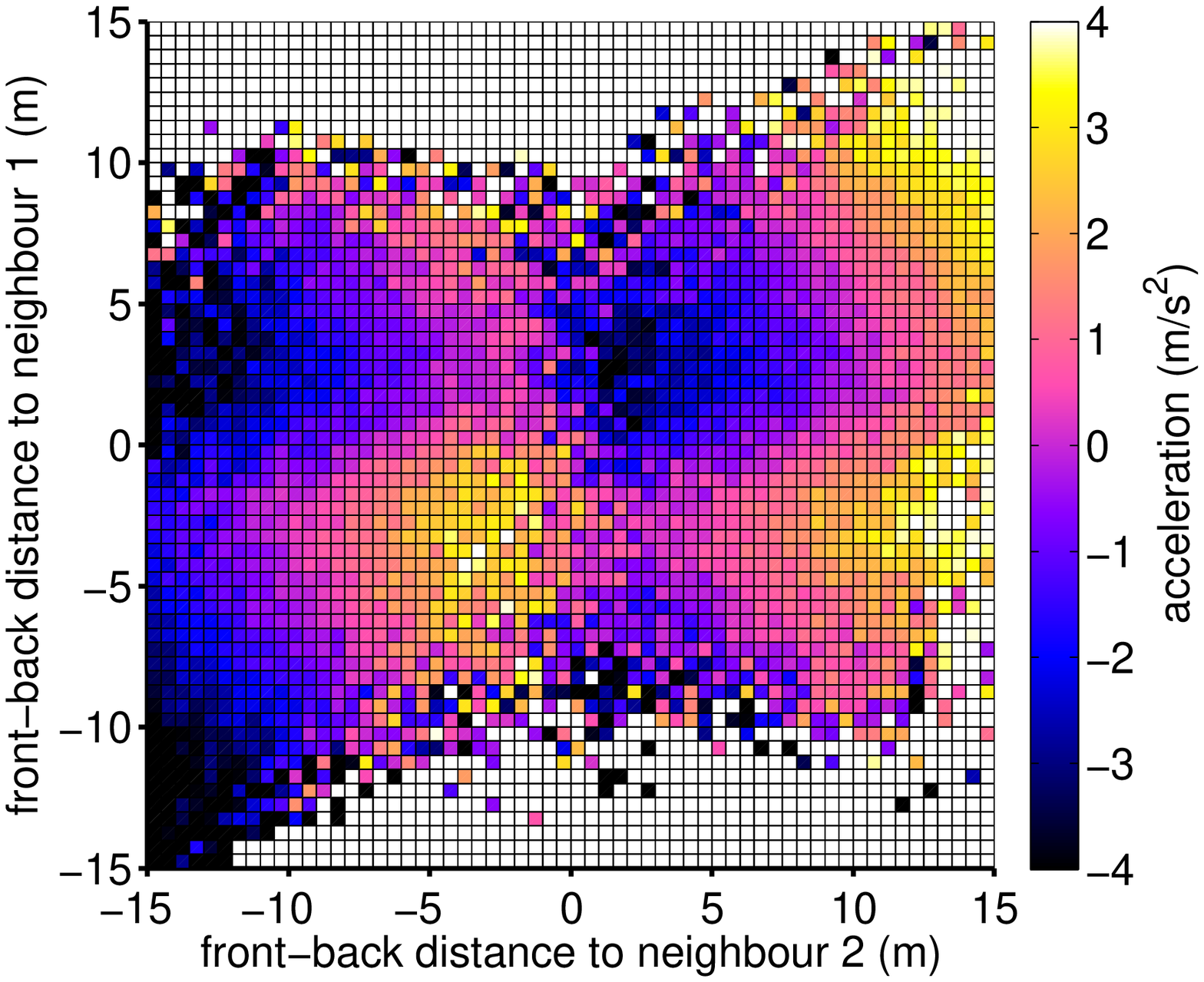} &
\includegraphics[width=0.45\textwidth]{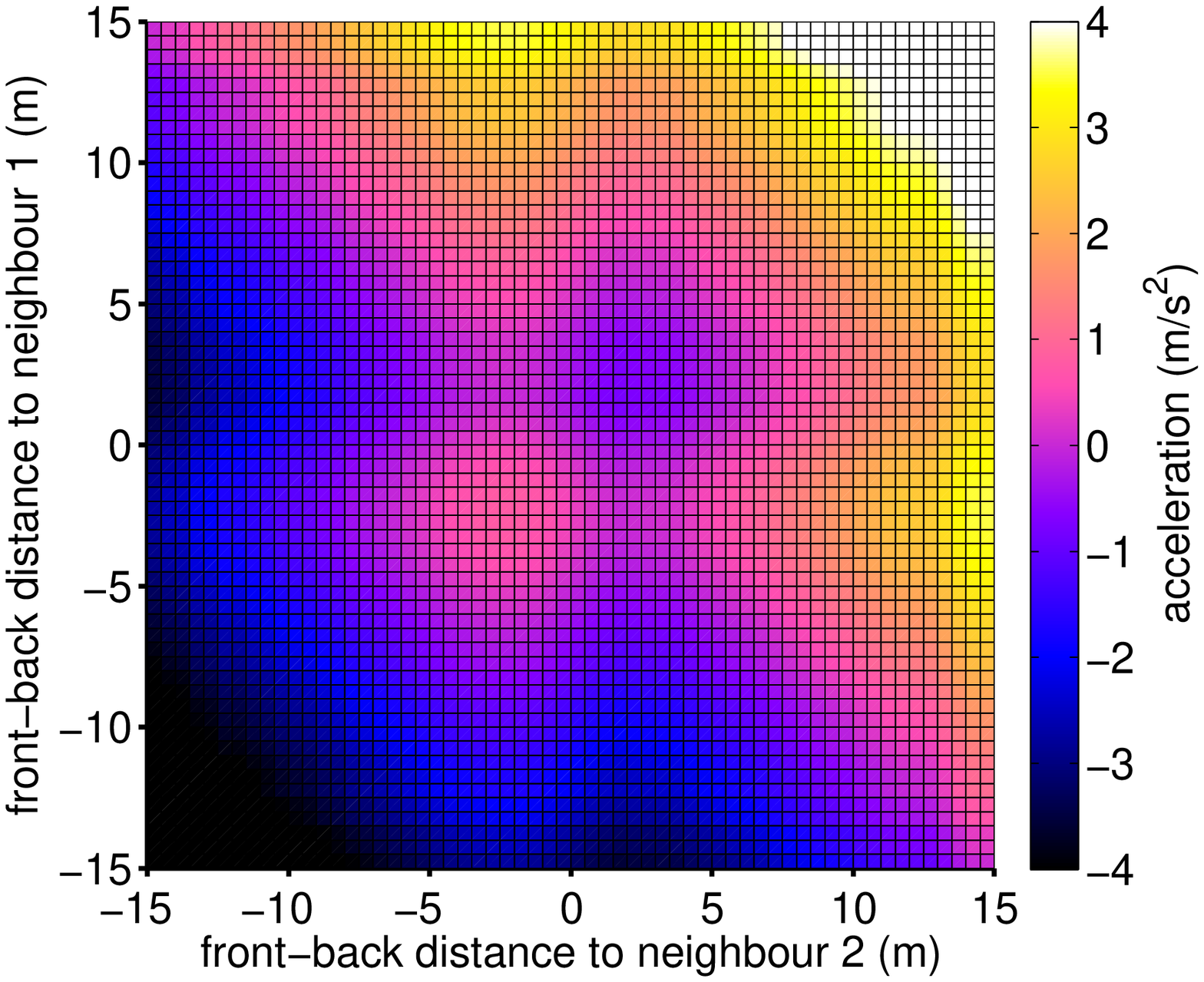} \\
\includegraphics[width=0.45\textwidth]{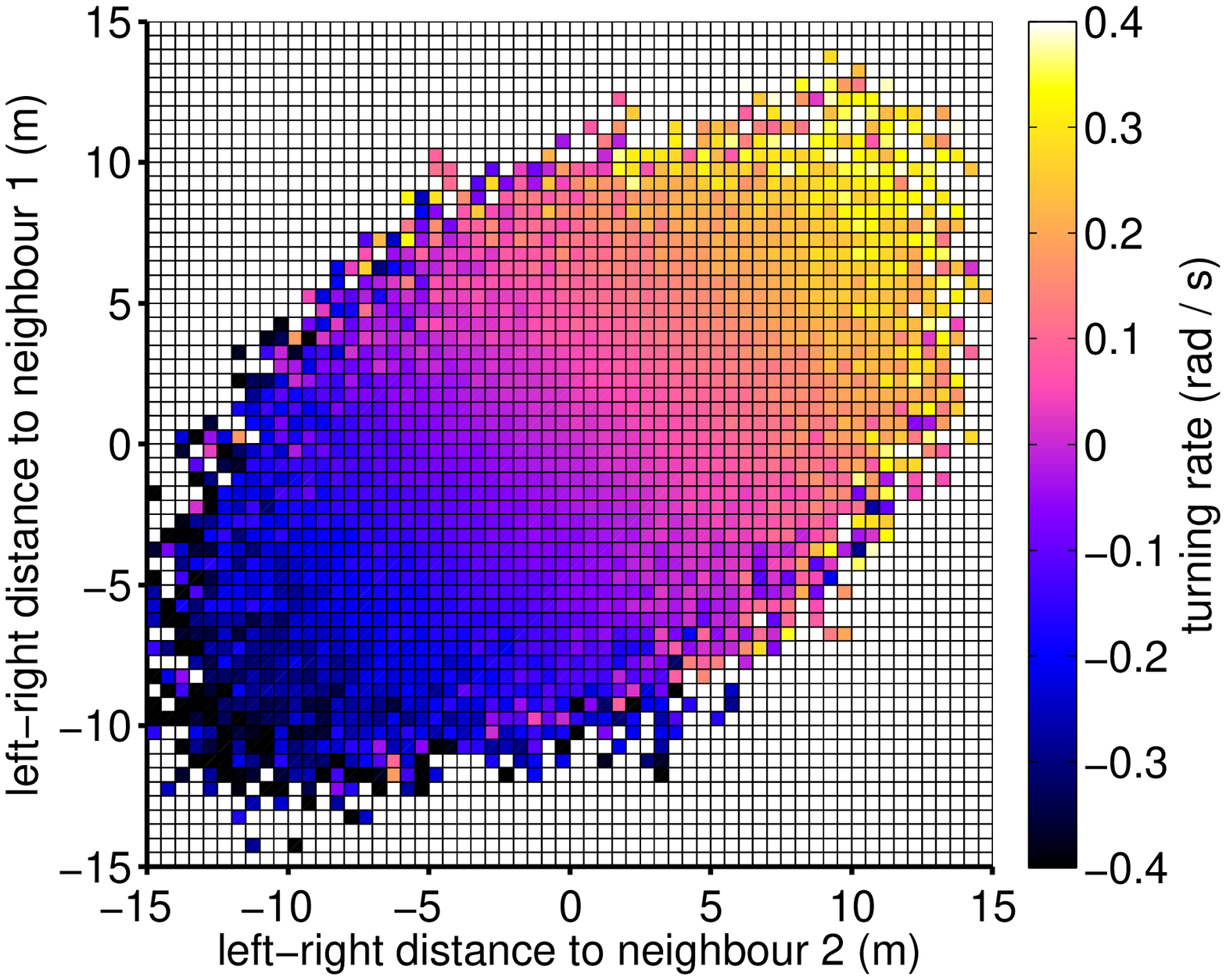} &
\includegraphics[width=0.45\textwidth]{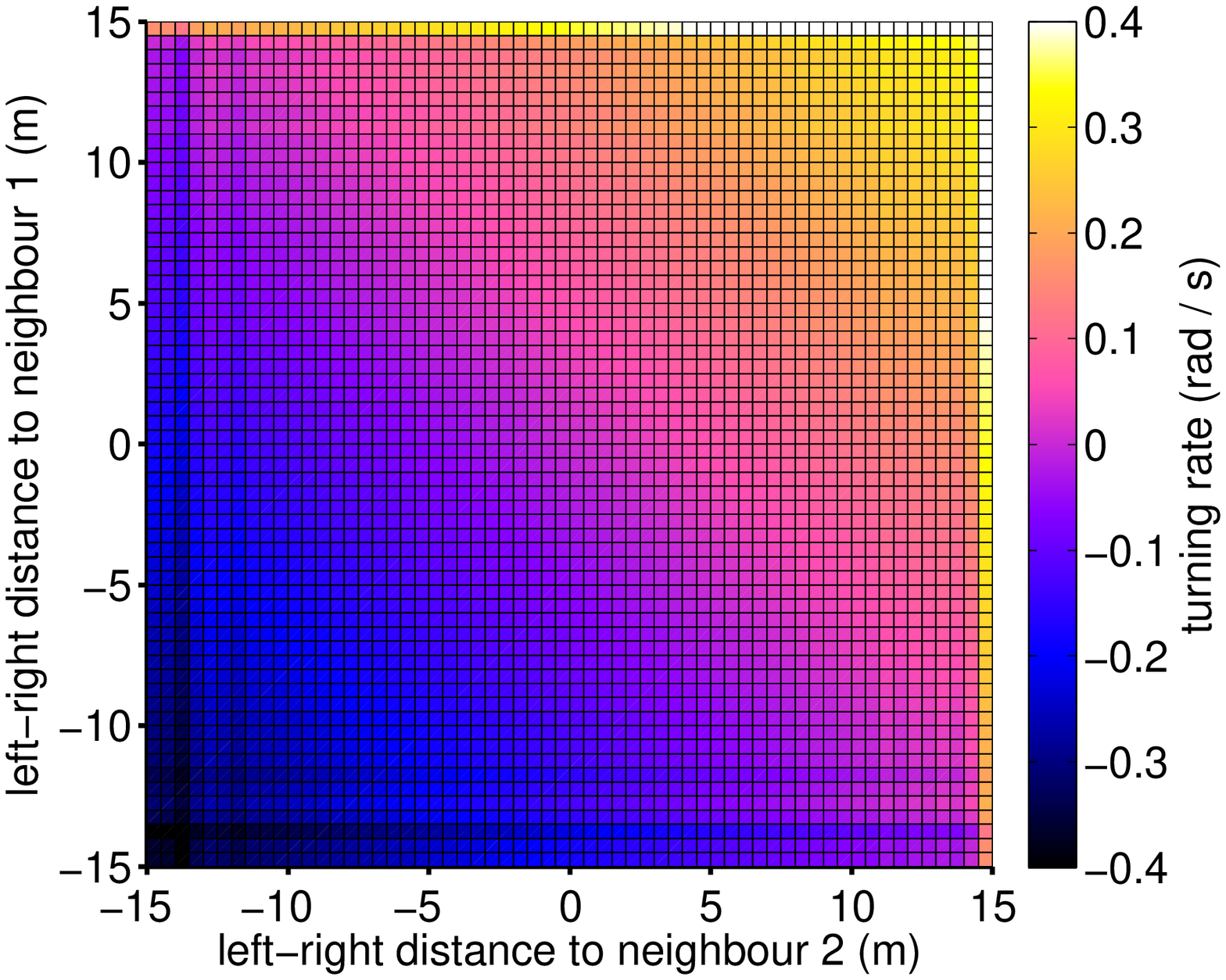}
\end{tabular}
\end{minipage}
\caption{
\csentence{Observed and predicted responses to multiple neighbours.} {\bf Top row} Observed (left) and predicted (right) acceleration response in groups of three individuals. {\bf Bottom row} Observed (left) and predicted (right) turning responses. Predicted responses are calculated by combining the observed responses in simulations with two individuals (one single neighbour) under the assumption that the combined effect of two neighbours is equal to the average of two independent pairwise responses. White squares in the grids on the left indicate missing values, never occurring in the simulations.
}
\label{fig:multiple_interacting_neighbours}
\end{figure}

%%%%%%%%%%%%%%%%%%%%%%%%%%%%%%%%%%%
%%                               %%
%% Additional Files              %%
%%                               %%
%%%%%%%%%%%%%%%%%%%%%%%%%%%%%%%%%%%

\section*{Additional Files}
  \subsection*{Additional file ``main.m''}
Matlab\textsuperscript{\textregistered} script file to run the analyses reported in this paper.
\subsection*{Other additional files}
 Matlab\textsuperscript{\textregistered} functions required by ``main.m''.

\end{backmatter}
\end{document}